\documentclass[11pt]{article}
\pdfoutput=1
\usepackage[margin=1.0in]{geometry}
\usepackage{amsmath,amssymb,graphicx}
\usepackage{hyperref}
\usepackage{slashed}

\usepackage[font=small,labelfont=bf]{caption}
\usepackage{cite}

\newcommand{\be}{\begin{equation}}
\newcommand{\ee}{\end{equation}}

\usepackage{xcolor}
\usepackage{bm}

\newcommand{\sigmabar}{{\overline \sigma}}

\title{\bf Last Electroweak WIMP Standing\\
{\large Pseudo-Dirac Higgsino Status and Compact Stars as Future Probes}}
\author{Rebecca Krall and Matthew Reece\\
{\small \em Department of Physics, Harvard University, Cambridge, MA, 02138}
}

\begin{document}
\maketitle

\begin{abstract}
Electroweak WIMPs are under intense scrutiny from direct detection, indirect detection, and collider experiments. Nonetheless the pure (pseudo-Dirac) higgsino, one of the simplest such WIMPs, remains elusive. We present an up-to-date assessment of current experimental constraints on neutralino dark matter. The strongest bound on pure higgsino dark matter currently may arise from AMS-02 measurements of antiprotons, though the interpretation of these results has sizable uncertainty. We discuss whether future astrophysical observations could offer novel ways to test higgsino dark matter, especially in the challenging regime with order MeV mass splitting between the two neutral higgsinos. We find that heating of white dwarfs by annihilation of higgsinos captured via inelastic scattering could be one useful probe, although it will require challenging observations of distant dwarf galaxies or a convincing case to be made for substantial dark matter content in $\omega$ Cen, a globular cluster that may be a remnant of a disrupted dwarf galaxy. White dwarfs and neutron stars give a target for astronomical observations that could eventually help to close the last, most difficult corner of parameter space for dark matter with weak interactions.
\end{abstract}

\section{Introduction}

One of the most pressing problems in particle physics is the nature of dark matter. Although abundant evidence from cosmology and astrophysics points to the existence of approximately cold, collisionless dark matter, we do not know anything else about the identity or interactions of dark matter. One compelling possibility is that dark matter is a thermal relic, i.e.~that it was once in chemical equilibrium with the Standard Model (SM) and froze out as the universe expanded. This idea has led to the WIMP paradigm, with dark matter as a Weakly Interacting Massive Particle that annihilated to Standard Model particles in the early universe with a rate $\langle \sigma v \rangle \approx 2 \times 10^{-26}~{\rm cm}^3/{\rm s}$ (see e.g.~\cite{Steigman:1984ac, Gondolo:1990dk, Steigman:2012nb}). This corresponds to cross sections roughly the size of those determined by the Standard Model weak interaction, a coincidence that is often called the ``WIMP miracle.''

The purest form of the WIMP miracle would arise if dark matter were made up of chiral particles interacting with the electroweak gauge group and obtaining a mass from the Higgs mechanism. In that case, both the mass and the interactions of the WIMP would be pinned to the weak scale. (The mass could be below the weak scale, but the simplest formulation would have an order-one Yukawa coupling.) However, such models have long been ruled out by data. Among the remaining models, we can distinguish non-chiral WIMP dark matter interacting through {\em the} Standard Model weak interaction from models that postulate more general interactions of the same approximate strength as, but differing from, the SM weak interaction. Neither of these models is completely miraculous, as the former requires further explanation of why the WIMP {\em mass} is near the weak scale, while the latter is completely unrelated to the Standard Model weak scale without further model-building. Non-chiral electroweak WIMPs have also been called ``Minimal Dark Matter'' \cite{Cirelli:2007xd}. Examples arise in supersymmetric theories, which automatically provide dark matter candidates in SU(2) doublets (higgsinos) and a triplet (wino) together with a singlet (bino) that can mix to provide small splittings in the multiplets via high-dimension operators.

In this paper we first provide an up-to-date review of the experimental status of neutralino dark matter, with an emphasis on the case of mostly-higgsino dark matter. We reinterpret the latest dark matter direct and indirect detection constraints and show that although wino dark matter is highly constrained, higgsino dark matter is largely unconstrained. The regime of higgsino parameter space in which the two neutral higgsinos have a quite small mass splitting, of order an MeV, is a difficult regime to experimentally probe but is well-motivated in the context of particular models including realizations of Spread SUSY \cite{Hall:2011jd} and Split Dirac SUSY \cite{Fox:2014moa}. Similar scenarios have also been advocated as the minimal model for dark matter and unification \cite{Mahbubani:2005pt} or as the minimal version of $Z$-mediated dark matter compatible with precision electroweak constraints \cite{Kearney:2016rng}. If the splitting were much smaller than an MeV, higgsinos could be easily ruled out via inelastic scattering of the light eigenstate to the heavy one via $Z$ boson exchange \cite{TuckerSmith:2001hy, TuckerSmith:2004jv}. However, this regime of much smaller splittings is theoretically implausible anyway, as it requires heavy gauginos that give rise to large threshold corrections lifting higgsinos well above a TeV. 

Inelastic dark matter may be probed by dark matter capture in compact stars, i.e.~white dwarfs and neutron stars, which have large escape velocities due to their very high densities. As a result, infalling dark matter acquires significant kinetic energy and can possibly upscatter to a heavier mass eigenstate. Observed neutron stars have radii of around 10 kilometers and masses of 1.17 to 2.0$M_\odot$ \cite{Antoniadis:2013pzd, Ozel:2016oaf}. Their equation of state is not yet known, because the composition of matter at high densities needs to be understood \cite{Ozel:2016oaf,Chamel:2008ca}. Given substantial uncertainties in the composition of neutron stars, we will largely focus on the case of white dwarfs, which are better understood and could also probe the regime of MeV mass splittings. From the mass distribution of observed white dwarfs, the most likely mass occurs around 0.5-0.7 $M_{\odot}$ \cite{Kepler:2006ns}. The typical radius of a white dwarf is about 1\% of the Sun's radius, or about 7000 km \cite{1979ApJ...228..240S}. They are formed when a star expands to a red giant and fuses helium to carbon and oxygen in its core. If the mass of the star isn't large enough to generate the core temperatures required to fuse carbon, an inert mass of carbon and oxygen builds up at the center. The outer layers shed to form a planetary nebula, and the core becomes the white dwarf \cite{Shapiro:1983du}. From a Monte Carlo simulation of stellar models the mean values of the central carbon-12 and oxygen-16 mass fractions are 33\% and 63\%, respectively. The scattering of dark matter on carbon and oxygen nuclei, including the appropriate nuclear form factors, is well-understood \cite{Catena:2015uha}. We will compute the capture rate of higgsino dark matter in white dwarf stars and explore whether observations could provide useful new constraints. The outlook is somewhat unclear, as achieving a sufficiently strong constraint will require either powerful new observations of distant regions of high dark matter density (like dwarf galaxies) or perhaps an argument that more nearby targets contain larger amounts of dark matter than may have been anticipated.

The outline of this paper is as follows. In \S\ref{sec:higgsino}, we will review the basic properties of higgsinos, with an emphasis on the physics of the small mass splittings and the couplings of the higgsino in the regime of small mixings with gauginos. In \S\ref{sec:directdetection}, we interpret the current direct detection results of experiments like PandaX-II as constraints on the parameter space of neutralino dark matter. We also present the expected future reach of experiments like Xenon1T and LZ. In \S\ref{sec:indirectdetection}, we interpret measurements of gamma rays by the Fermi-LAT and HESS telescopes and of antiprotons by AMS-02 as constraints on neutralino parameter space. In \S\ref{sec:otherconstraints}, we briefly summarize other constraints (current and anticipated) from colliders and searches for electric dipole moments, then recap the overall picture, which is that higgsinos that are well-mixed with gauginos are easily constrained by several probes but that the regime of small mixing is quite difficult to access. Next, in \S\ref{sec:capture}, we consider the capture of higgsinos through inelastic scattering in compact stars, i.e.~white dwarfs and neutron stars. We concentrate on the case of white dwarfs, where measurements of the temperature distribution of large numbers of white dwarfs could reveal a floor from dark matter annihilations in the star. Again, this method of constraining higgsino dark matter proves challenging, and will require a future generation of telescopes. We highlight that the unknown dark matter distribution in Omega Centauri, a globular cluster believed to be a tidally disrupted dwarf galaxy, could provide a relatively nearby target. This possibility deserves further study. We offer some concluding remarks in \S\ref{sec:conclusions}, including a back-of-the-envelope check that monochromatic MeV gamma rays from ${\widetilde H}^0_1 {\widetilde H}^0_1 \to {\widetilde H}^0_2 {\widetilde H}^0_2$ have too low a rate to be likely to be observed by upcoming satellites like e-ASTROGAM.

\section{Properties of higgsinos}
\label{sec:higgsino}

\begin{figure}[!h]\begin{center}
\includegraphics[width=1.0\textwidth]{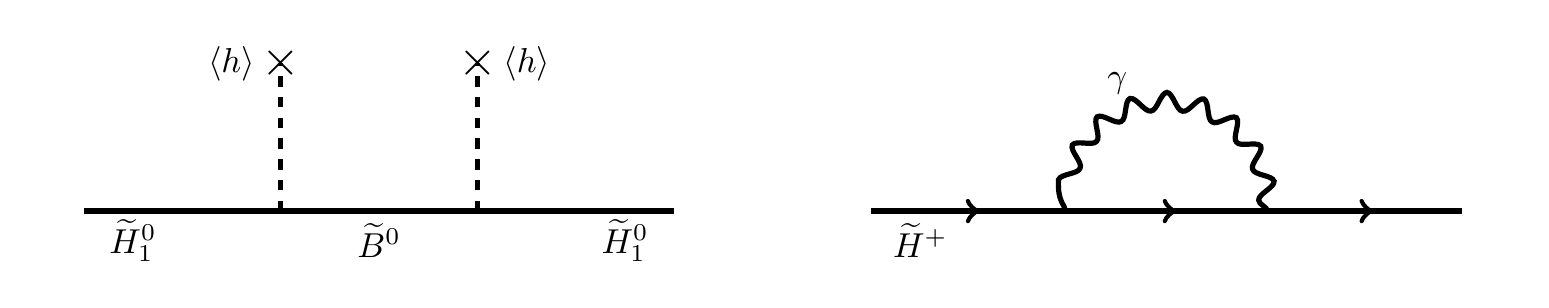}
\end{center}
\caption{Effects that generate a mass splitting between the higgsino states. Left: tree-level splittings among both charged and neutral states arise due to mixings with gauginos and are of order $m_Z^2/M_{1,2}$. Right: loop-level splittings add about 350 MeV to the chargino mass while leaving the two neutralinos degenerate \cite{Thomas:1998wy}.}
\label{fig:higgsinosplittingdiagrams}
\end{figure}

In this section we will review some basic properties of higgsinos, emphasizing the origin of small mass splittings and the physics arising when the splittings are small. Higgsinos are a pair of doublet fermions, ${\widetilde H}_u$ and ${\widetilde H}_d$, linked by a Dirac mass $\mu {\widetilde H}_u \cdot {\widetilde H}_d$. Before electroweak breaking, this leads to a neutral Dirac fermion and a charged Dirac fermion with identical masses. After electroweak symmetry breaking, these are split into two neutral Majorana fermions together with a charged Dirac fermion. The neutral fermions are split by tree level effects related to mixing with gauginos and the chargino is split by similar tree-level effects and also one-loop effects that are independent of the gaugino masses, as illustrated in Fig.~\ref{fig:higgsinosplittingdiagrams}. In the MSSM, the mass splitting between the two neutral Majorana higgsinos, which we will denote $\delta$ to connect to the literature on inelastic dark matter, is approximately
\be
\delta \approx m_Z^2 \left(\frac{\sin^2 \theta_W}{M_1} + \frac{\cos^2 \theta_W}{M_2}\right).
\ee
This result assumes that the higgsino mass parameter $\mu$ is real, and neglects some running effects on the gaugino--higgsino--gauge boson couplings; a more general result may be found in \cite{Nagata:2014wma}. Furthermore, in the context of models where one or more of the gauginos have Dirac masses, the result may be further suppressed. For instance, in the ``Hypercharge Impure'' model of \cite{Fox:2014moa}, effectively the $M_1$ term is present but the $M_2$ term is much smaller, because the mass of the wino is dominantly a Dirac mass. The chargino is split by tree-level effects of similar size, but in the large $M_1$, $M_2$ limit, it is dominantly made heavier by a one-loop effect that raises its mass by about 350 MeV \cite{Thomas:1998wy, Cirelli:2005uq, Buckley:2009kv}.

If neutralinos are to be thermal relic WIMPs, we have a few options: for generic weak-scale masses we can fine-tune a mixture of gauginos and higgsinos to obtain a ``well-tempered neutralino'' \cite{ArkaniHamed:2006mb}; we can have a pure higgsino at about 1.1 TeV or a pure wino at around 3 TeV \cite{Hisano:2006nn}; or we can have pure higgsinos or winos that are lighter but acquire a larger relic abundance through a non-thermal mechanism like a late-decaying gravitino or modulus \cite{Moroi:1999zb, Acharya:2009zt}. A higgsino of mass near 1 TeV is a good candidate to keep in mind for much of the discussion in this paper.

The limit of very pure higgsinos occurs when the bino and wino are much heavier (for more detailed exploration of models in this limit, see for instance \cite{Hall:2011jd,Fox:2014moa}). In that limit, if the bino and wino masses are dominantly Majorana (as in the MSSM), there is a threshold correction to the higgsino mass:
\be
\Delta \mu \approx -\frac{\sin 2\beta}{32\pi^2} \left(3 g^2 M_2 \log\frac{m_{H}}{M_2} + g'^2 M_1 \log \frac{m_H}{M_1}\right),
\ee
with $m_H$ the mass scale of the heavy higgs bosons. Without fine-tuning we expect that the low energy value satisfies $|\mu| \gtrsim |\delta \mu|$. For dark matter, we are chiefly interested in the regime $|\mu| \lesssim 1.1~{\rm TeV}$, which then leads to a requirement that $M_{1,2}$ are not too large, e.g.~$M_1 \lesssim 3 \times 10^6~{\rm GeV}$. This, in turn, leads to a minimum expectation for the mass splitting between the two neutralino states. For example, if we focus on the bino (as motivated by the Hypercharge Impure model of \cite{Fox:2014moa}, assuming the wino decouples via a Dirac mass), we have
\begin{align}
\delta & \gtrsim g'^2 \sin 2\beta \sin^2 \theta_W \frac{m_Z^2}{32 \pi^2 \mu} \log\frac{m_H}{M_1} \\
&\approx 970~{\rm keV} \frac{1.1~{\rm TeV}}{|\mu|} \frac{\log(m_H/M_1)}{\log(100)} \frac{\sin(2\beta)}{\sin(2\arctan(2))} \nonumber \\
&\approx 240~{\rm keV} \frac{1.1~{\rm TeV}}{|\mu|} \frac{\log(m_H/M_1)}{\log(100)} \frac{\sin(2\beta)}{\sin(2\arctan(10))}
\end{align}
where in the last two lines we have chosen a modest value of the logarithm and shown results for the cases $\tan \beta = 2$ and $\tan \beta = 10$. The conclusion is that without fine-tuning to maintain $|\mu|$ more than a loop factor below $M_{1,2}$, we tend to expect that the tree-level mass splitting in the higgsino sector is at least a few hundred keV at large $\tan \beta$ or at least an MeV at small $\tan \beta$. Splittings below 200 keV or so are also in tension with the absence of inelastic dark matter scattering signals in direct detection experiments \cite{TuckerSmith:2001hy, Nagata:2014wma, Bramante:2016rdh}. On the other hand, as emphasized in \cite{Fox:2014moa}, the roughly MeV minimum splittings can fit nicely into an appealing scenario that maintains gauge coupling unification and achieves the right Higgs mass (though with the usual split SUSY fine-tuning to achieve the correct Higgs vev).

\begin{figure}[!h]\begin{center}
\includegraphics[width=1.0 \textwidth]{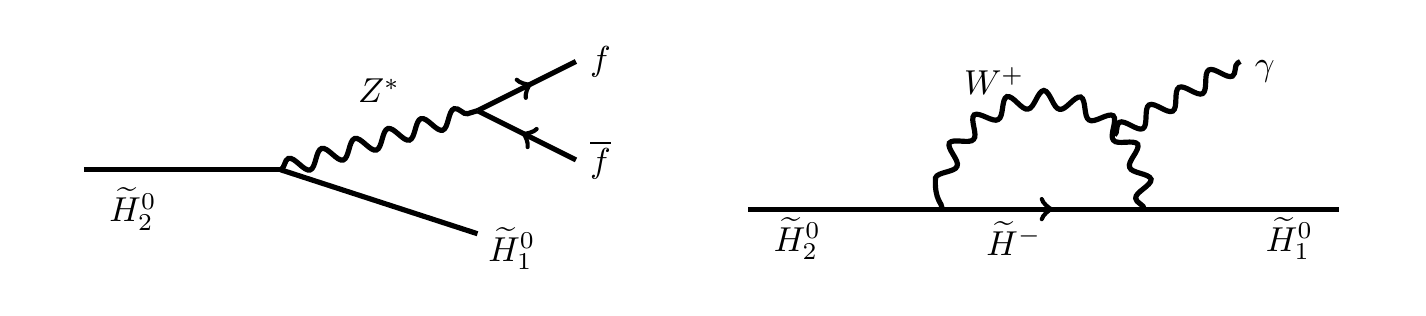}
\end{center}
\caption{Decays of the heavier neutral higgsino mass eigenstate to the lighter neutral higgsino mass eigenstate.}
\label{fig:higgsinodecaydiagrams}
\end{figure}%

In the nearly-degenerate regime in which the mass of ${\widetilde H}^0_2$ is between those of ${\widetilde H}^0_1$ and ${\widetilde H}^\pm_1$, the heavier neutral higgsino decays to the light neutral higgsino plus either two fermions or one photon, as illustrated in Fig.~\ref{fig:higgsinodecaydiagrams}. The tree-level decay width ${\widetilde H}^0_2 \to {\widetilde H}^0_1 {\overline f} f$ is given by 
\begin{align}
\Gamma({\widetilde H}^0_2 \to {\widetilde H}^0_1 {\overline f} f) = C_{\rm tree} \frac{g^4 \delta^5}{480 \pi^3 m_Z^4 \cos^4 \theta_W},
\end{align}
with $C_{\rm tree}$ a constant depending on the $Z$ boson couplings to species $f$. In particular, we have the following values depending on which states are summed (assuming the mass splitting is well above the threshold for each):
\begin{align}
C_{\rm tree} &\approx \begin{cases} 
                             0.75, & \nu_e, \nu_\mu, \nu_\tau \\
                             0.88 & \nu_e, \nu_\mu, \nu_\tau, e \\
                             3.7 & \nu_e, \nu_\mu, \nu_\tau, e, \mu, \tau, u, d, s, c, b, \\
                          \end{cases}
\end{align} 
so although the precise width is difficult to calculate for mass splittings in the neighborhood of the QCD transition, the coefficient $C_{\rm tree}$ is always an order-one number. The radiative process ${\widetilde H}^0_2 \to {\widetilde H}^0_1 \gamma$ has a rate \cite{Haber:1988px}
\begin{align}
\Gamma({\widetilde H}^0_2 \to {\widetilde H}^0_1 \gamma) = C^2(\mu) \frac{e^2 g^4 \delta^3}{256 \pi^5 \mu^2} &\approx 6.4 \times 10^{-22}~{\rm GeV}  \left(\frac{\delta}{1~{\rm MeV}}\right)^3 \left(\frac{1~{\rm TeV}}{\mu}\right)^2 \frac{C^2(\mu)}{C^2({\rm TeV})} \nonumber \\
&\approx \frac{1}{3.1 \times 10^7~{\rm cm}}  \left(\frac{\delta}{1~{\rm MeV}}\right)^3 \left(\frac{1~{\rm TeV}}{\mu}\right)^2 \frac{C^2(\mu)}{C^2({\rm TeV})}.
\label{eq:decaywidth}
\end{align}
with $C^2(\mu)$ a loop function that depends on $\mu^2/m_W^2$ and is approximately given by $3.2 (\mu/{\rm TeV}) - 0.3$ over the range of higgsino masses between 200 GeV and 1 TeV. We find that the radiative decay to photons dominates over the tree level three-body decay ${\widetilde H}^0_2 \to {\widetilde H}^0_1 {\overline f} f$ whenever $\delta \lesssim 1~{\rm GeV}$. For later reference, we note that the lifetime $3 \times 10^7~{\rm cm}/c$ is three orders of magnitude smaller than the time it takes a particle moving at a speed $10^{-2} c$ to cross a white dwarf star.

Notice that although the lifetime of ${\widetilde H}^0_2$ is rather long, the emitted photon carries a small amount of energy. We do not expect such decays in the early universe to have any significant effect on BBN.

The existence of this radiative decay suggests the interesting possibility of observing monochromatic gamma rays of energy $\delta$ from regions with a sizable population of the excited state ${\widetilde H}^0_2$. We will return to this point below, though we have not found any plausible candidate source for detecting such a gamma ray line.

\section{Direct Detection Constraints on Neutralino Dark Matter}
\label{sec:directdetection}

The higgsino couples, through its mixings with gauginos, to the higgs boson, and so this coupling is suppressed by the ratio $m_Z/M_{1,2}$. As a result, probes of neutralino and chargino couplings to the higgs, including spin-independent direct detection and effects on higgs branching ratios, become quite small in the pseudo-Dirac limit. Furthermore, in this limit the $Z$ boson dominantly couples off-diagonally, to one ${\widetilde H}^0_1$ and one ${\widetilde H}^0_2$. This suppresses spin-dependent direct detection as well. As a result, in the small-splitting regime, higgsinos become extremely difficult to detect directly through their tree-level couplings. See, for example, \cite{Essig:2007az,Cohen:2010gj,Cheung:2012qy} for further discussion.

Beyond tree level, there remain one-loop processes through which pure higgsinos can scatter elastically in direct detection experiments. However, due to a cancelation between two diagrams that happens to be especially effective for the actual value of the higgs boson mass, the direct detection rate for higgsino dark matter via higgs exchange in the pure higgsino limit is tiny and in the neighborhood of the so-called ``neutrino floor'' at which direct detection experiments are no longer background-free \cite{Essig:2007az, Hisano:2011cs, Hisano:2012wm, Hill:2013hoa, Hill:2014yka, Hill:2014yxa}.

\begin{figure}[!h]\begin{center}
\includegraphics[width=\textwidth]{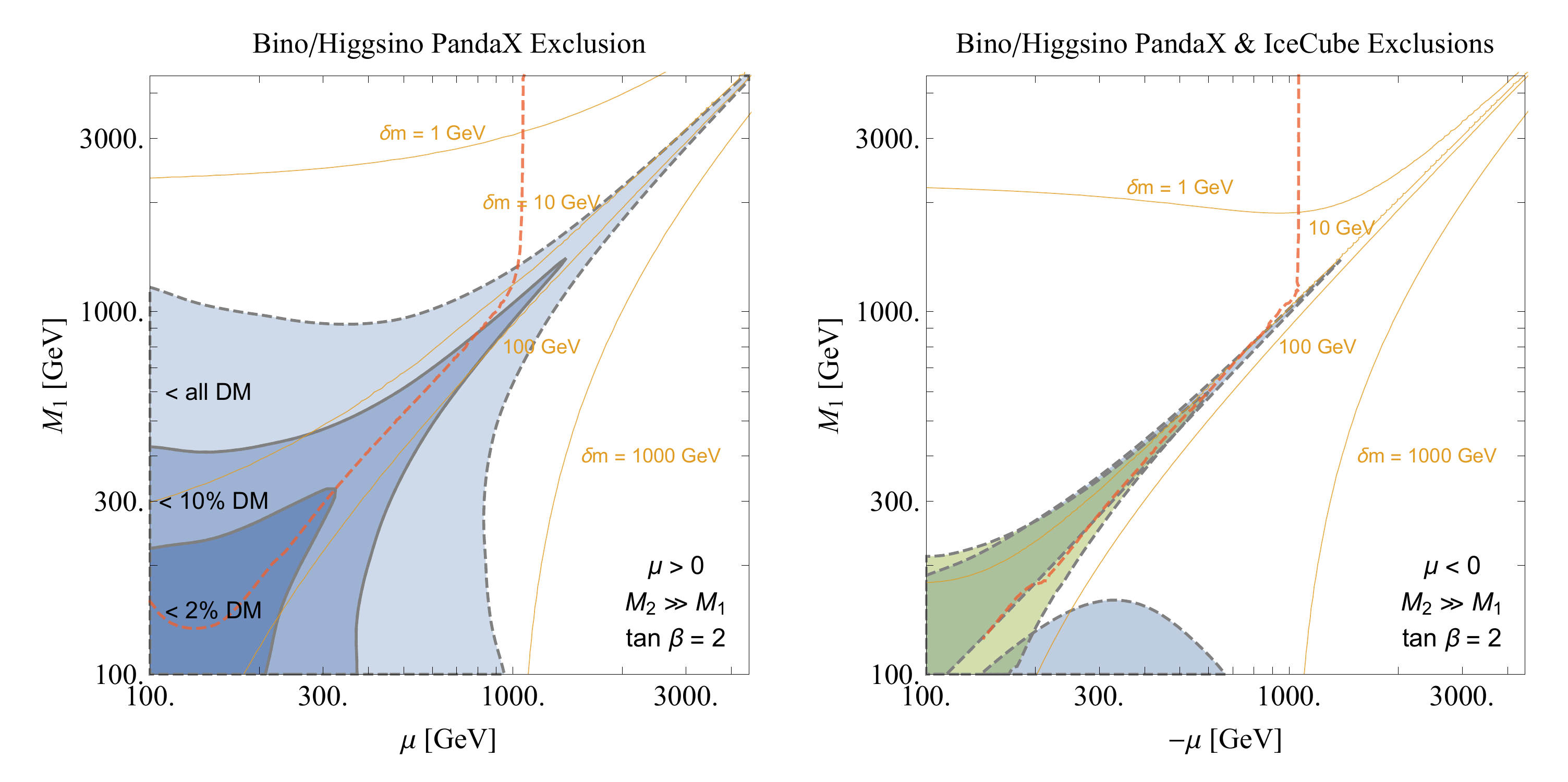}
\end{center}
\caption{Left: Direct detection constraints from the most up-to-date results of the PandaX experiment \cite{Cui:2017nnn}, for the case of mixed bino/higgsino dark matter with $\mu > 0$ and $\tan \beta = 2$. Right: PandaX constraints (blue) together with IceCube constraints (green) \cite{Aartsen:2016zhm} in the case with $\mu < 0$. The PandaX bound is from spin-independent scattering via higgs boson exchange; the IceCube constraint from spin-dependent scattering on protons in the Sun followed by annihilation. At left, because the bounds are quite strong, we also show constraints on the dark matter fraction. In both plots the thin orange curves show the mass difference between the two lightest neutralino mass eigenstates and the red dashed curve is where the thermal relic density $\Omega h^2 = 0.12$, as computed with micrOMEGAs \cite{Belanger:2013oya}.}
\label{fig:directdetectionlux}
\end{figure}%

Currently the strongest constraints come from the PandaX-II experiment's 2017 results \cite{Cui:2017nnn}. These set bounds slightly stronger than Xenon1T \cite{Aprile:2017iyp}, which even with a small fraction of its eventual data has surpassed the LUX \cite{Akerib:2016vxi} and 2016 PandaX \cite{Tan:2016zwf} bounds. Another constraint, which strictly speaking is not direct detection but which we include here due to the similar physics, arises from dark matter capture in the Sun \cite{Press:1985ug, Silk:1985ax, Gould:1987ir, Gould:1991hx}. Recently this has led to stringent constraints on many dark matter models based on IceCube searches for neutrinos originating in dark matter annihilation inside the Sun \cite{Aartsen:2016exj, Aartsen:2016zhm}. In the context of mixed bino/higgsino dark matter, this constraint is most relevant when $\mu < 0$. We show the current PandaX and IceCube constraints in Figure \ref{fig:directdetectionlux}. We present the case $\tan \beta = 2$ where the differences between $\mu > 0$ and $\mu < 0$ are starker than at large $\tan \beta$. (For further discussion of the relevant dependence on $\tan \beta$ and the sign of $\mu$, see \cite{Cohen:2010gj,Cheung:2012qy,Agrawal:2017zwh}.) Both the spin-independent coupling probed by PandaX and the spin-dependent coupling probed by IceCube rely on mixings with gauginos and hence the strongest constraints lie along the diagonal. Relatedly, from the orange contours in the figures we see that the regions being probed have large mass splittings between the two lightest neutralinos, which results from the large mixing. We also see from the red curves that if we insist on a thermal relic abundance, the bulk of mixed gaugino/higgsino parameter space is ruled out while the pure higgsino limit survives. In the case $\mu > 0$, the constraints are so strong that we find it useful to show regions where even a small fraction of neutralino dark matter is excluded. In the case $\mu < 0$, the constraints are relatively mild. To plot the IceCube constraint, we have used the official IceCube bound on dark matter annihilating to $W^+ W^-$ and have reweighted the effective spin-dependent cross section based on the mix of $W^+ W^-$, $ZZ$, and $t \overline{t}$ decays obtained according to micrOMEGAs \cite{Belanger:2013oya}; this procedure is similar to that used in \cite{Cheung:2012qy}.

\begin{figure}[!h]\begin{center}
\includegraphics[width=0.6 \textwidth]{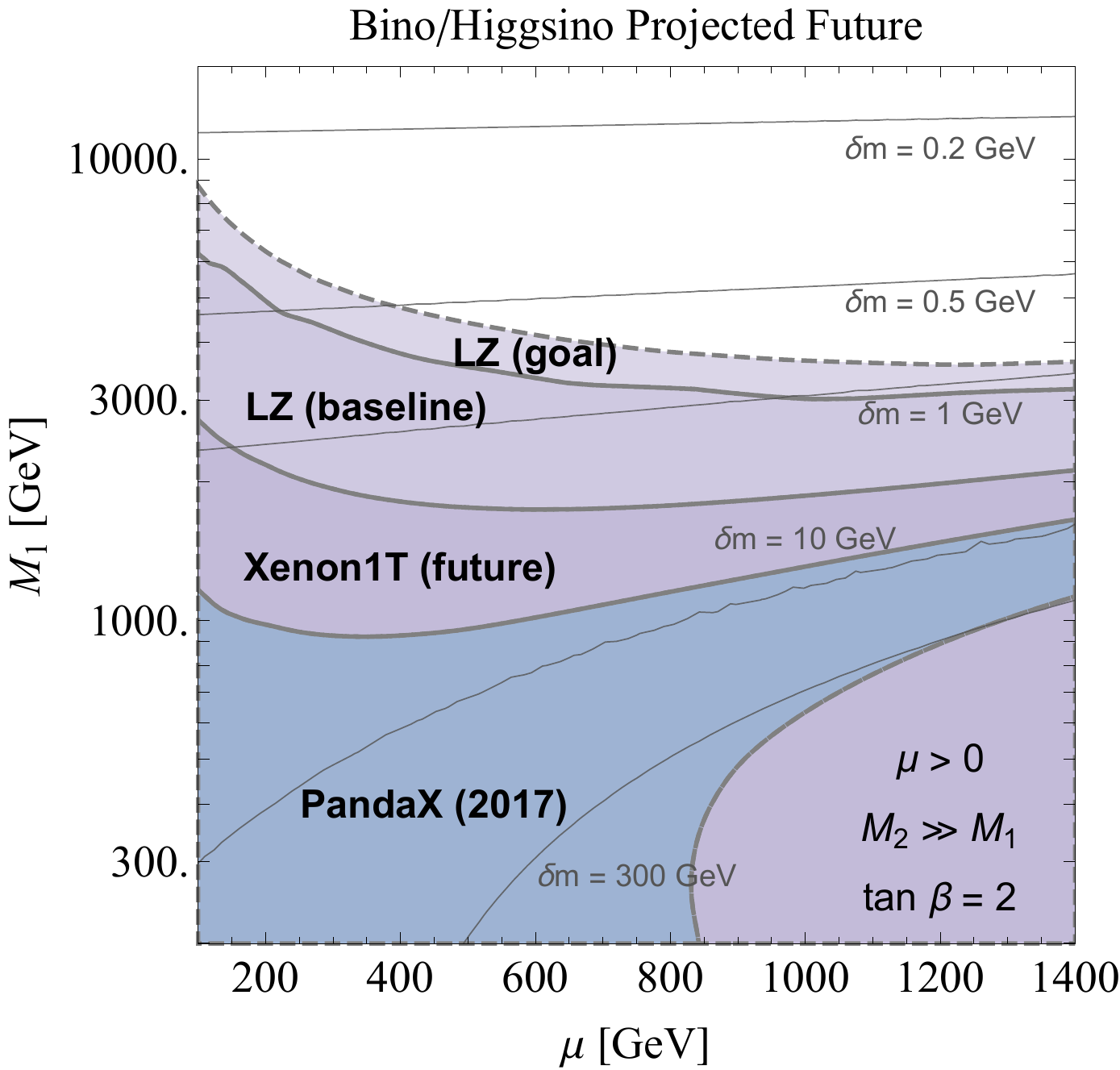}
\end{center}
\caption{Projected future direct detection constraints on mixed bino/higgsino dark matter. The shaded blue region is the current PandaX-II result (same as the left panel of Fig.~\ref{fig:directdetectionlux}). The shaded purple regions are projections for Xenon1T and LZ (with two different levels of optimism about what LZ will achieve). The thin gray curves show the mass splitting between the two lightest neutral higgsinos.}
\label{fig:directdetectionfuture}
\end{figure}%

In figure \ref{fig:directdetectionfuture} we include future projections from Xenon1T \cite{Aprile:2015uzo} and LZ (LUX-ZEPLIN) including both its baseline scenario and more optimistic goal \cite{Mount:2017qzi}. We see that the future reach of ton-scale direct detection experiments will probe gaugino masses of several TeV, as emphasized in \cite{Bramante:2015una}. However, they stop well short of probing the regime of PeV gauginos that arise, for example, in Split Dirac SUSY. We see that LZ will probe neutralino mass splittings $\delta$ of order or slightly below 1 GeV, but will not reach splittings significantly smaller than 1 GeV.

The current status of detection of higgsino dark matter through {\em inelastic} scattering has been recently discussed in \cite{Nagata:2014wma, Bramante:2016rdh}. Roughly speaking, direct detection experiments rule out the range of mass splittings $\delta < 200~{\rm keV}$, including the pure Dirac doublet or ``heavy neutrino'' dark matter candidate that was excluded many years ago. Searches for events over a wider range of recoil energies could potentially probe mass splittings as large as $\delta \approx 0.5~{\rm MeV}$ \cite{Bramante:2016rdh}, though this depends crucially on the velocity distribution of dark matter in our neighborhood. A recent attempt to empirically determine this velocity distribution by using old stars as tracers of the dark matter halo points to rather lower than expected velocity dispersion \cite{Herzog-Arbeitman:2017fte}. If the inferred distribution proves to be accurate, the reach of terrestrial experiments for inelastic dark matter would be significantly reduced. 

\section{Indirect Detection Constraints on Neutralino Dark Matter}
\label{sec:indirectdetection}

Data from indirect detection can place significant constraints on any dark matter candidate in an SU(2)$_L$ multiplet, due to the annihilation $\chi^0 \chi^0 \to W^+ W^-$ (and $ZZ$, depending on the representation), which leads to copious production of gamma rays and antiprotons \cite{Cirelli:2007xd}. These searches are effective even for very small mass splittings within the SU(2)$_L$ multiplet, and thus are complementary to direct detection.  Here we will present the current constraints on a few scenarios on pure wino and pure higgsino dark matter. For the computation of wino annihilation, the Sommerfeld enhancement is crucial \cite{Hisano:2004ds}, and the one loop annihilation process $\chi^0 \chi^0 \to \gamma \gamma$ may also be detectable \cite{Bergstrom:1997fh, Bern:1997ng}. As a result, in recent years sophisticated effective field theory techniques have been applied to more accurate computation of these annihilation processes \cite{Hryczuk:2011vi, Bauer:2014ula, Ovanesyan:2014fwa, Baumgart:2014saa, Baumgart:2015bpa, Ovanesyan:2016vkk}, especially in the case of gamma ray line searches. 

\begin{figure}[!h]\begin{center}
\includegraphics[width=0.55\textwidth]{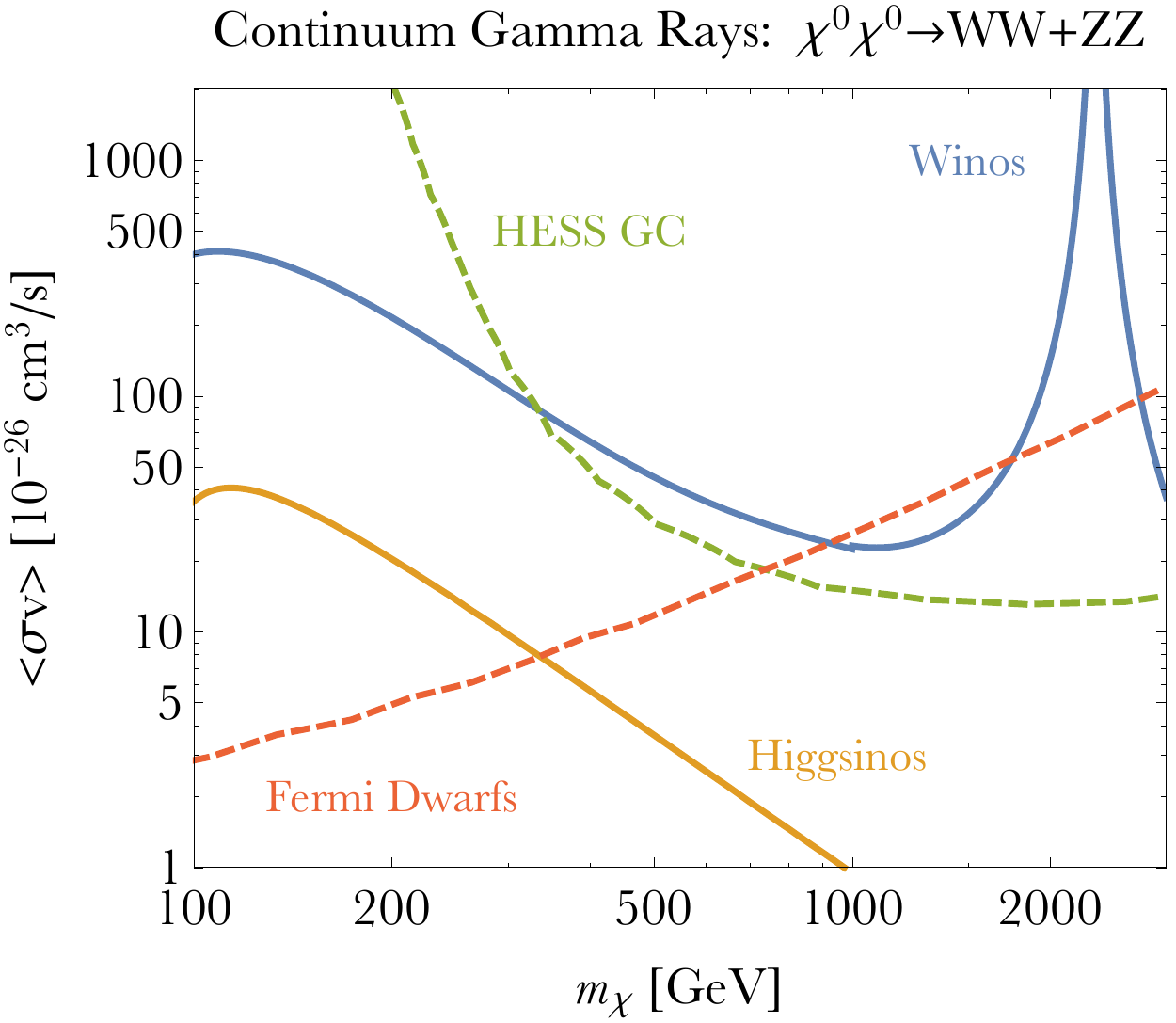} 
\end{center}
\caption{Indirect detection constraints on wino (blue) and higgsino (orange) dark matter from continuum gamma ray spectra. The region above the dashed lines is excluded by Fermi-LAT observations of gamma rays from dwarf galaxies \cite{Ackermann:2015zua} and HESS observations of the galactic center \cite{Abdallah:2016ygi}. The Sommerfeld effect is included in the wino calculation \cite{Hisano:2004ds}.}
\label{fig:higgsinoindirectdetection}
\end{figure}%

Gamma ray constraints arise from Fermi-LAT searches for photons from dwarf galaxies \cite{Ackermann:2015zua} (which set the most stringent constraints at low masses) and HESS observations of the galactic center \cite{Abdallah:2016ygi} (which are more effective at high masses). We summarize the current constraints in Figure \ref{fig:higgsinoindirectdetection}. Notice that we have plotted data only from continuum gamma ray signals. Winos are excluded as the dominant component of dark matter over the full mass range, while higgsinos have smaller annihilation rates and largely escape from the constraints. Gamma ray line searches with HESS have previously been argued to exclude high mass wino dark matter \cite{Cohen:2013ama}; we find that their continuum search has now achieved sufficient sensitivity as well. In the case of the galactic center, there are large astrophysical uncertainties, and the bounds could be ameliorated if our galaxy's dark matter halo has a kiloparsec-size core \cite{Fan:2013faa}. Fermi-LAT's dwarf galaxy bound already marginalizes over uncertainties in the $J$-factors arising from the unknown distribution of dark matter. We find that this bound excludes the possibility that higgsinos constitute all of the dark matter up to masses of about 330 GeV. In the computation of higgsino annihilation we have used micrOMEGAs \cite{Belanger:2013oya}. We have treated the $W^+W^-$ and $ZZ$ final states as equivalent in terms of the photons and antiprotons they yield; this is not exact, but is approximately true. One could also include the tree-level internal bremsstrahlung process \cite{Bergstrom:2005ss}, but using this in a full likelihood fit of Fermi-LAT data does not dramatically change the answer.\footnote{Personal communication from Wei Xue.}

The most recent release of AMS-02 data on antiprotons \cite{Aguilar:2016kjl} has been used to place stringent limits on dark matter annihilation to hadronic final states \cite{Cuoco:2016eej, Cui:2016ppb}. As recently noted in \cite{Kawamura:2017amp}, the central exclusion curve of \cite{Cuoco:2016eej} (and the similar curve of \cite{Cui:2016ppb}) excludes higgsino dark matter up to masses of about 800 GeV. However, substantial systematic uncertainties exist in the interpretation of antiproton data, including among others the unknown amount of convection in cosmic ray propagation, the height of the diffusion zone, and uncertainties in antiproton production cross sections \cite{Korsmeier:2016kha}. The conservative end of the systematic uncertainty band of \cite{Cuoco:2016eej} excludes higgsino dark matter up to masses of about 480 GeV, while its optimistic end reaches above 930 GeV. 

\begin{figure}[!h]\begin{center}
\includegraphics[width=\textwidth]{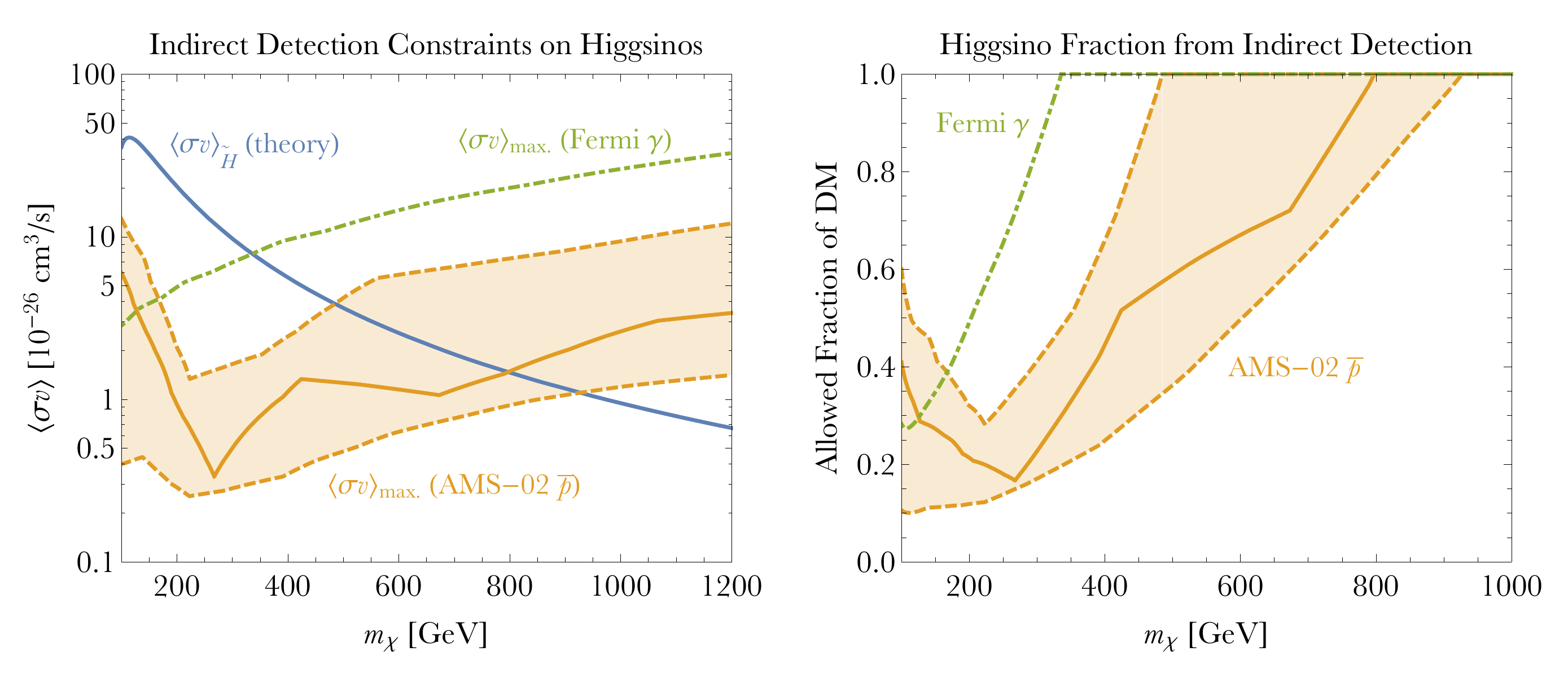} 
\end{center}
\caption{Indirect detection constraints on higgsino dark matter. Left: predicted averaged annihilation cross section to $W^+ W^-$ and $ZZ$ (if higgsinos are all of the dark matter) compared to experimental constraints from Fermi-LAT observations of gamma rays from dwarf galaxies \cite{Ackermann:2015zua} and AMS-02 measurements of antiprotons \cite{Aguilar:2016kjl} as interpreted by Cuoco et al.~\cite{Cuoco:2016eej}. The shaded orange region represents the systematic uncertainty band quoted in \cite{Cuoco:2016eej}.}
\label{fig:higgsinoindirectdetection}
\end{figure}%

To summarize, although wino dark matter is under severe stress from indirect detection data, higgsino dark matter is relatively safe. Light higgsinos, below around 500 GeV, are in tension with antiproton data, while gamma ray data covers even less ground. Even the most strongly constrained regions of higgsino parameter space still allow the higgsino to constitute about one-third of all the dark matter in the universe. While it is possible that a significant reduction in uncertainties about cosmic ray propagation could shrink the error bar associated with the AMS-02 exclusion, at the moment it seems prudent to attach a very large amount of uncertainty to the result. Similarly, while further observations of stellar kinematics in dwarf galaxies could reduce uncertainties in $J$-factors, this is unlikely to dramatically alter the interpretation of Fermi-LAT data. More significant improvements might come with future gamma-ray telescopes. Still, future telescopes like the Cherenkov Telescope Array (CTA) will struggle to cover the full higgsino dark matter parameter space \cite{Buckley:2013bha, Cahill-Rowley:2013dpa, Fan:2013faa, Cahill-Rowley:2014boa}. For these reasons, it is especially interesting to search for new ways to cover portions of the higgsino parameter space that are otherwise out of experimental reach.

Nonetheless, given the current results, a careful analysis of the present and future status of indirect detection of higgsino dark matter would be timely. Resummation will play an important role in precisely determining the rate of gamma-ray lines from higgsino dark matter annihilation, especially for larger masses near 1 TeV, where until further theoretical progress is made it remains unclear whether CTA will place a constraint \cite{Baumgart:2015bpa}.

\section{Other constraints and summary of higgsino status}
\label{sec:otherconstraints}

\subsection{Electron EDM}

\begin{figure}[!h]\begin{center}
\includegraphics[width=0.5\textwidth]{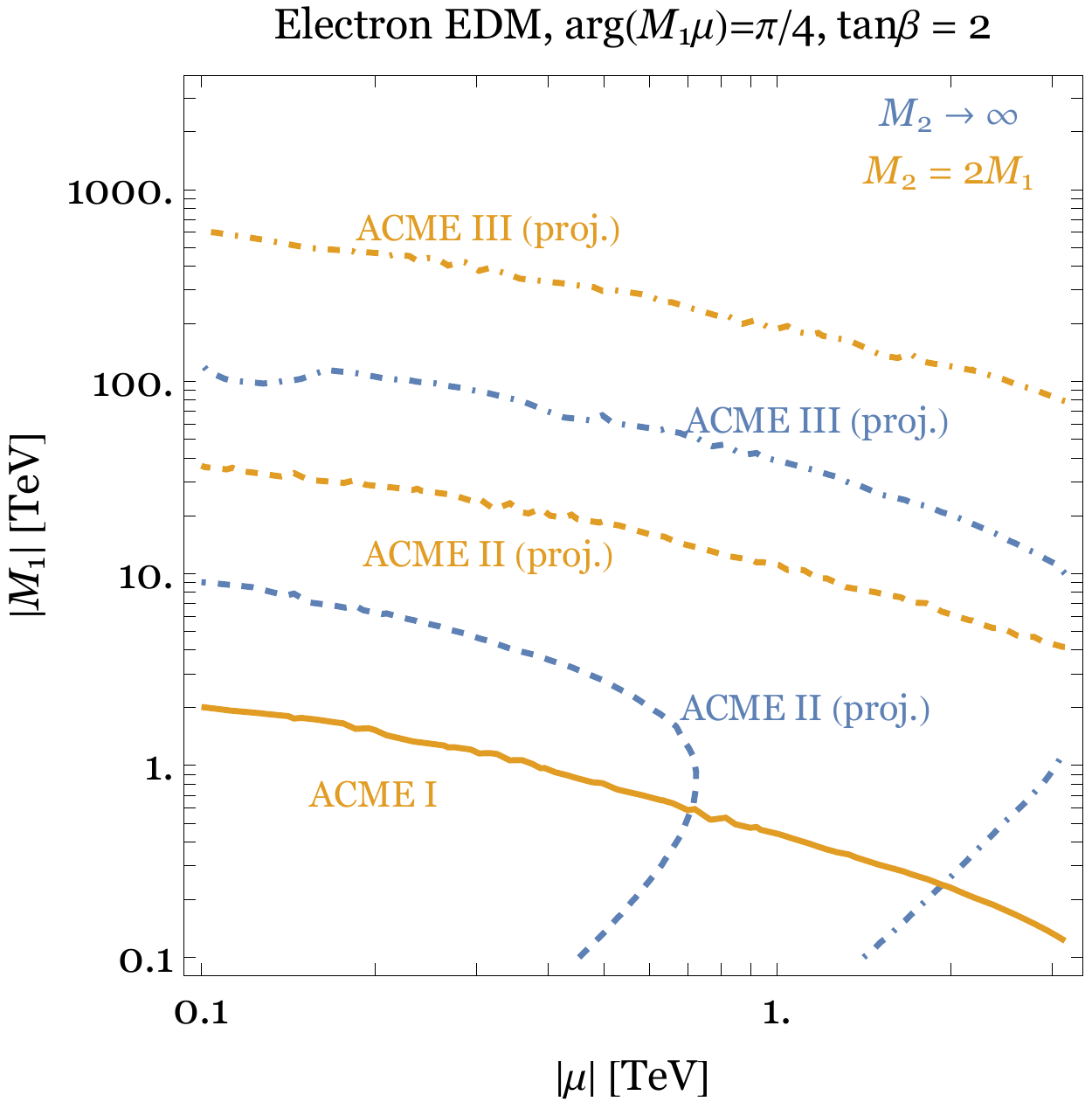}
\end{center}
\caption{Current and future reach of the ACME search for an electron EDM. We have fixed a large CP-violating phase, ${\rm arg}(M_1 \mu) = \pi/4$. We show two cases: in orange, the wino mass is fixed to twice the bino mass. In blue, we decouple the wino while keeping the bino in the spectrum, as in the Hypercharge Impure model of \cite{Fox:2014moa}. We see that this case is much less constrained.}
\label{fig:EDMplot}
\end{figure}%

Charginos and neutralinos induce an electric dipole moment of the electron at two loops from Barr-Zee type diagrams \cite{Barr:1990vd}, which has been extensively discussed \cite{Pilaftsis:2002fe, ArkaniHamed:2004yi, Giudice:2005rz, Li:2008kz}. The current strongest constrain on the electron EDM comes from the ACME I experiment \cite{Baron:2013eja}, which will be substantially improved in the near future with results from ACME II and ACME III \cite{Panda:2016ifg,Doyletalk}. A recent analysis of current and near-future electron EDM constraints on charginos appears in \cite{Nakai:2016atk}. In Figure \ref{fig:EDMplot}, we have shown the expected future constraints in the region of parameter space where the gaugino masses are much larger than the higgsino mass, assuming that there is an order-one CP-violating phase. We see that the ACME III projected reach ($|d_e| \lesssim 0.3 \times 10^{-30} e~{\rm cm}$) extends to the several hundred TeV regime of gaugino masses, where the neutralino mass splitting $\delta$ is of order tens of MeV. It does not extend all the way to the MeV splitting regime. The bound is notably stronger when the winos are not decoupled, because then the diagrams connecting a chargino loop to the electron line with $\gamma h$ and $Z h$ lines contribute. In the case with only a bino in the spectrum, the leading effects connect a chargino-neutralino loop to the electron line with $W W$ lines and are significantly smaller.

\subsection{Collider searches}

\begin{figure}[!h]\begin{center}
\includegraphics[width=0.5\textwidth]{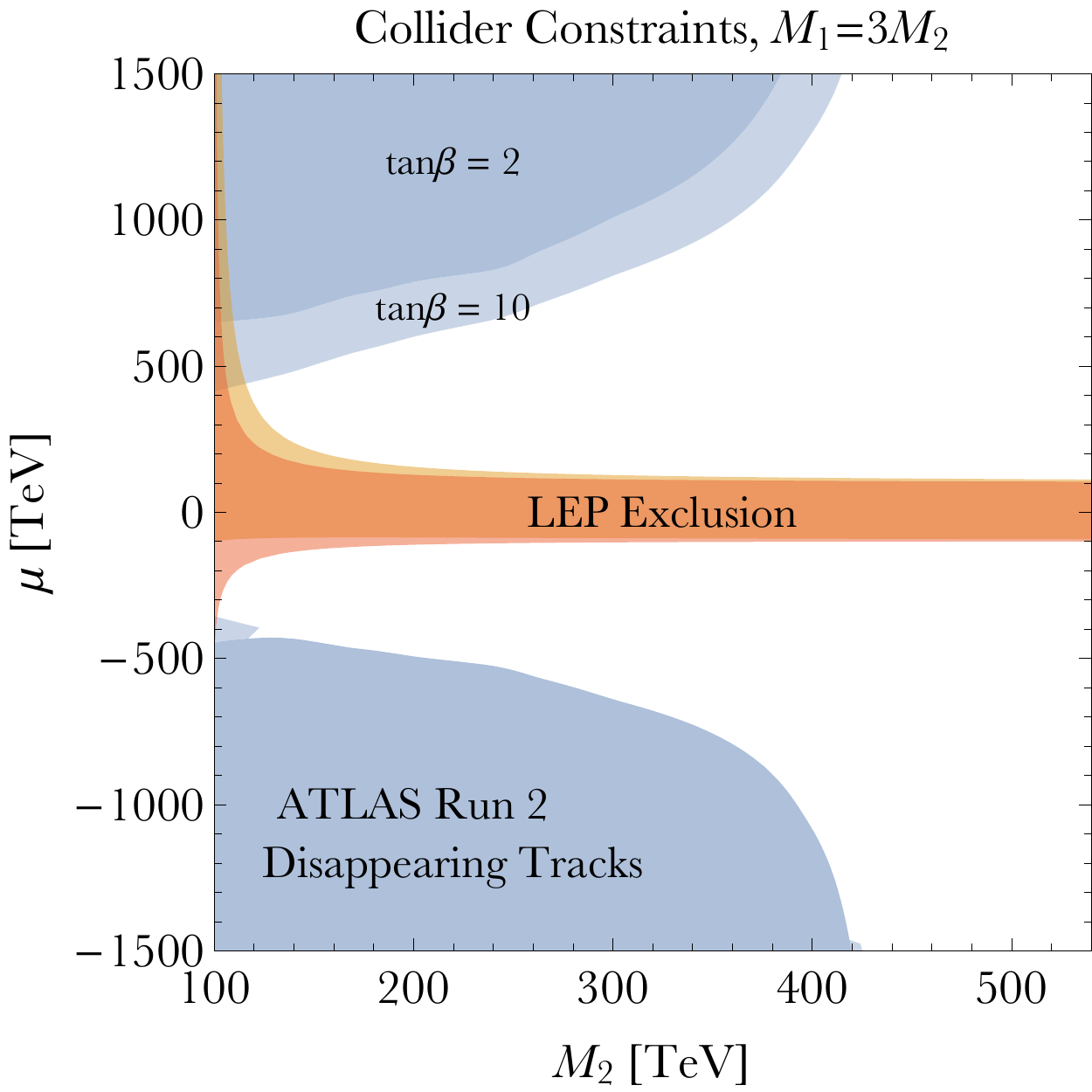}
\end{center}
\caption{Results of the recent ATLAS search for disappearing tracks \cite{ATLAS:2017bna}, reinterpreted in a scenario where the wino is not a pure triplet but mixes with higgsinos and binos.}
\label{fig:charginocolliderplot}
\end{figure}

SUSY searches at the LHC have placed stringent constraints on the production of gluinos and squarks, but have had relatively little impact so far on our knowledge of electroweakinos, at least without assuming light sleptons to produce dramatic lepton-rich signatures. One of the more significant results arises from a search for disappearing tracks, which arise when a chargino propagates a macroscopic distance before decaying to a neutralino and a very soft charged particle that is unobserved \cite{Chen:1995yu,Chen:1999yf}. The recent update from ATLAS excludes nearly degenerate winos up to 430 GeV \cite{ATLAS:2017bna}. When the wino mixes with other superpartners, the bounds can become substantially weaker, as illustrated in Fig.~\ref{fig:charginocolliderplot}. (This updates a similar plot in \cite{Nakai:2016atk}; see also related results in \cite{Han:2016qtc}.) Although existing disappearing track searches have set interesting bounds on winos, they do not yet constrain higgsinos, for two reasons: the higgsino lifetime is typically shorter (so that the chargino usually does not pass through many layers of the tracker before decaying, unless it is highly boosted) and the higgsino cross section is also substantially smaller.

If other superpartners are decoupled, higgsinos alone can be a very difficult target for colliders. Simple monojet searches at the LHC will only marginally improve on the results of LEP, and even at a future higher energy hadron collider, the higgsino is a challenging target \cite{Halverson:2014nwa, Low:2014cba, Ismail:2016zby}. A linear collider, if one is constructed and achieves energies of a TeV or higher, can directly probe a large fraction of higgsino parameter space, although a version of CLIC with center of mass energy above 2.2 TeV would be needed to reach the thermal relic higgsino. Indirect probes at $e^+ e^-$ colliders also will cover only a portion of the parameter space \cite{Harigaya:2015yaa, Wu:2017kgr}. Very recently, there has been new theoretical work on how optimized searches for disappearing tracks could constrain higgsinos \cite{Mahbubani:2017gjh,Fukuda:2017jmk} (related work appeared earlier in \cite{Halverson:2014nwa}). In the pure higgsino limit where the chargino-neutralino mass difference arises at loop level, they project that the LHC can exclude higgsinos up to around $400$ to $600~{\rm GeV}$ and a future 33 TeV or 100 TeV hadron collider could probe the full range up to a thermal relic higgsino at 1.1 TeV. While these results add to the appeal of a future hadron collider, it lies some decades in the future, and in any case would never tell us if the higgsinos constitute a significant fraction of dark matter. From one point of view, this is a virtue: higgsinos could be discovered even if they are a highly subdominant component of dark matter or are unstable on long timescales. However, it does leave a strong motivation for experimental probes that could test not just the existence of higgsinos but whether they are a substantial component of dark matter.

\subsection{Summary: status of higgsino dark matter}

Dark matter composed dominantly of a higgsino is a difficult challenge for both dark matter searches and collider searches. Direct detection and EDMs are powerful probes of the regime with large mixings with gauginos, and as EDM experiments achieve increasing sophistication they will even probe the regime where gaugino masses are in the 100s of TeV. However, the reach of EDM experiments crucially depends on CP-violating phases, which may be small in some models. The most powerful and universal probe of dark matter in the unmixed or small mass splitting regime is indirect detection, which looks for ${\widetilde H}^0_1 {\widetilde H}^0_1 \to W^+ W^-, ZZ$ and hence relies only on the gauge interactions of the higgsino. The bounds from looking for such annihilations in gamma rays from dwarf galaxies are fairly robust but not very strong; even in the range of masses that they exclude, they allow higgsinos to be more than 10\% of the dark matter. Bounds from antiprotons are potentially stronger but subject to large uncertainties, and will require further study to make a truly convincing case that we understand the astrophysics well enough to be confident about the particle physics conclusions. All of this leaves us with a gap in coverage, especially at large higgsino masses near the 1.1 TeV thermal relic and at small mass splittings in the MeV range. These regimes of parameter space are interesting, being motivated by various models and theoretical arguments \cite{Mahbubani:2005pt, Hall:2011jd, Fox:2014moa}. In the next section we will discuss a challenging but not yet fully explored avenue for understanding this part of parameter space, and make the case for further work in astrophysics to determine how promising it is.

\section{Capture of higgsinos in compact stars}
\label{sec:capture}

In this section we will consider the capture of higgsino dark matter in compact stars. We focus on white dwarfs, although many of our conclusions should carry over to neutron stars (with rather different observational prospects). Capture of dark matter in compact stars has been discussed in the literature \cite{Goldman:1989nd, Kouvaris:2007ay, Bertone:2007ae, Fan:2011dw, Bertoni:2013bsa, Hurst:2014uda}. One major novelty with such objects, compared to the Sun, is kinematic: because they are very dense they have a high escape velocity. The escape velocity at the surface of the Sun is about 600 km/s, not dramatically larger than the faster WIMPs in the galactic halo. Although inelastic scattering of dark matter in the Sun is interesting, it probes similar mass splittings to those probed by direct detection on Earth \cite{Nussinov:2009ft, Menon:2009qj, Shu:2010ta}. Compact stars potentially probe the parameter space at larger mass splittings. If we estimate the escape velocity at the surface of a white dwarf to be $2 \times 10^{-2} c$, then using a carbon-12 nucleus as our target (which determines the reduced mass) we would kinematically expect to probe splittings up to about 2 MeV for a 1 TeV WIMP. A similar estimate for neutron stars indicates that they could probe splittings up to several hundred MeV. In the interior of the stars the escape velocity can be even larger. This makes white dwarfs and neutron stars important potential probes of inelastic dark matter: they are capable of capturing dark matter that would be kinematically unable to scatter on Earth or in the Sun. This dark matter can accumulate inside the star; its subsequent annihilations would then heat the star. Because white dwarfs and neutron stars do not burn nuclear fuel, they are expected to gradually cool over the age of the universe. Heating via dark matter annihilations could prevent such cooling from proceeding indefinitely. In some regions of the galaxy a floor on white dwarf temperatures has been observed; comparing such floors to the minimum amount of heating by dark matter could produce a constraint on, or an indirect signal of, inelastic dark matter annihilation \cite{McCullough:2010ai, Hooper:2010es}.

As this paper was being completed, \cite{Baryakhtar:2017dbj} appeared which discusses capture of dark matter, including higgsinos, in neutron stars. They assess the observational prospects for future detection of very cold neutron stars. This appears to be a challenging, but perhaps feasible, observation.

\subsection{Couplings for inelastic DM scattering}

The dark matter coupling to the $Z$ boson has the form \cite{Essig:2007az}
\be
\frac{ig}{2c_W} Z_\mu \left({\widetilde H}^\dagger_2 \sigmabar^\mu {\widetilde H}_1 - {\widetilde H}_1^\dagger \sigmabar^\mu {\widetilde H}_2\right).
\ee
The largest component of inelastic scattering goes through the vector coupling of the $Z$ boson to quarks, i.e.~to the combination of charges $\frac{g}{2 c_W} (T_3 - 2 s_W^2 Q)$ where $T_3 = 1/2, -1/2$ for the up and down quarks respectively. Because the vector current is conserved, we can simply add up the couplings to quarks to obtain the coupling to nucleons:
\be
\frac{g}{4 c_W} Z_\mu \left[\left(1 - 4 s_W^2\right) {\overline p} \gamma^\mu p - {\overline n} \gamma^\mu n\right].
\ee
Because $s_W^2 \approx 1/4$, the coupling to protons is much smaller than the coupling to neutrons. For this reason, at low momentum transfer we approximately expect the dark matter to scatter coherently off the neutrons, with a cross section scaling as $N^2$ where $N = A-Z$ is the number of neutrons.

\begin{figure}[!h]\begin{center}
\includegraphics[width=0.45\textwidth]{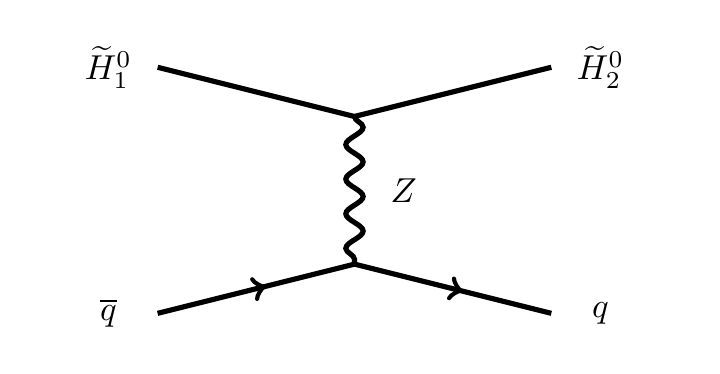}
\end{center}
\caption{The inelastic scattering process of a higgsino on a nucleus, ${\widetilde H}^0_1 N \to {\widetilde H}^0_2 N$.}
\label{fig:inelastic}
\end{figure}%

If we, for the moment, neglect the nuclear form factor and the effect of the mass splitting, the dark matter--nucleus scattering cross section is \cite{Nagata:2014wma,Bramante:2016rdh}
\be
\sigma_{\chi N} = \frac{G_F^2 \mu_{\chi N}^2}{8 \pi} \left[N - (1 - 4 s_W^2) Z\right]^2,
\ee
with $\mu_{\chi N}$ the dark matter--nucleus reduced mass. Evaluating for carbon-12 nuclei with mass 11.2 GeV and $N = Z = 6$, this is
\begin{align}
\left. \sigma_{\chi N} \right|_{m_\chi = 1000~{\rm GeV}} & \approx 2.8 \times 10^{-8}~{\rm GeV}^{-2} \approx 1.1 \times 10^{-35}~{\rm cm}^2. \label{eq:sigmachiN}
\end{align}
Of course, the form factors will impose some cost, but this is a large cross section, so when inelastic scattering is kinematically accessible the rates can be very large.

In our detailed calculations below we use nuclear form factors for carbon and oxygen found in \cite{Catena:2015uha}, which bases its calculation on \cite{brown2014shell, Anand:2013yka}. Because higgsinos have a large vector-current coupling, form factors for subleading operators have little effect on the result.

\subsection{Capture rate in a white dwarf: rare scattering limit}
\label{sec:particlecapturerate}

The computation of the capture rate of inelastic dark matter in a star, generalizing the classic calculation of Gould \cite{Gould:1987ir}, has been discussed in \cite{Nussinov:2009ft, Menon:2009qj, Shu:2010ta}. The appendix of \cite{Nussinov:2009ft} explaining this capture rate calculation contains a small mistake (though \cite{Shu:2010ta} uses similar notation and is correct), so let us briefly review the correct procedure.\footnote{We thank Prateek Agrawal for dusting off old notes to confirm our understanding of how to reconcile the various formulations in the literature.} We can parametrize the kinematics of scattering in terms of the kinetic energy $Q$ imparted to the nucleus in the rest frame of the incoming nucleus. The kinetic energy of the outgoing dark matter particle is lower than that of the incoming dark matter particle by an amount $Q + \delta$, because an energy $\delta$ is absorbed to excite the heavier state. If the incoming dark matter speed at the collision point is $w = \sqrt{u^2 + v(r)^2}$, with $u$ the speed at infinity and $v(r)$ the escape velocity at radius $r$ in the star, the kinematically accessible values of $Q$ lie between $Q_{\rm min}$ and $Q_{\rm max}$ where
\be
Q_{\rm min, \rm max} = \frac{1}{2} m_\chi w^2 \left[1 - \frac{\mu^2}{m_N^2} \left(1 \pm \frac{m_N}{m_\chi}\sqrt{1 - \frac{\delta}{\mu w^2/2}}\right)^2\right] - \delta.
\ee
We can consider the outgoing dark matter particle to be captured when
\be
Q > Q_{\rm cap} = \frac{1}{2} m_\chi (w^2 - v(r)^2) - \delta.
\ee
In other words, $Q_{\rm cap}$ is the threshold at which the {\em outgoing} dark matter particle has kinetic energy $\frac{1}{2} m_\chi v(r)^2$ and would marginally escape from the star. Given the formulas above, it can happen that the computed value of $Q_{\rm cap}$ is smaller than $Q_{\rm min}$ for some choices of velocity at infinity $u$, and so one must take care to integrate over nucleus energies not from $Q_{\rm cap}$ to $Q_{\rm max}$ but rather from $Q_{\rm min}' = {\rm max}(Q_{\rm cap}, Q_{\rm min})$ to $Q_{\rm max}$.

The capture rate in a volume element $dV$ is given by 
\be
\frac{dC}{dV} = \int_0^\infty du\, \frac{f(u)}{u} w \Omega(w),
\ee
where $\Omega(w)$ is the rate per unit time at which a WIMP of velocity $w$ will scatter to a velocity less than $v(r)$. It is given by
\be
\Omega(w) = \sum_i n_i(r) \int_{Q_{\rm min}'}^{Q_{\rm max}} dE\, \frac{d\sigma_i}{dE}(w^2, E).
\ee
Here $\frac{d\sigma_i}{dE}$ is the differential cross section for dark matter scattering by nuclei of mass $m_i$ and stellar density $n_i(r)$. We integrate it over the nuclear recoil energy. The total capture rate is then given by
\be
C = 4\pi \int^{R_*}_0 dr\, r^2 \frac{dC}{dV}.
\ee
In all of our calculations, we approximate the distribution of dark matter velocities at infinity with a Maxwell-Boltzmann distribution with velocity dispersion $v_0$ as seen by an observer with speed $v_*$,
\be
f(x) dx = \frac{\rho_\chi}{m_\chi} \frac{4}{\sqrt{\pi}} x^2 e^{-x^2} e^{-\eta^2} \frac{\sinh(2 x \eta)}{2 x \eta} dx,
\ee
with $x$ and $\eta$ dimensionless quantities
\be
x^2 \equiv \frac{u^2}{v_0^2}, \quad \eta^2 \equiv \frac{v_*^2}{v_0^2}.
\ee
We expect that in general $v_* \sim v_0$, that is, the typical speed of the star in which we capture is of order the velocity dispersion in the halo, so $\eta \sim 1$. We find that the quantitative impact of varying $\eta$ from 0 to 1 is fairly small. 

\begin{figure}[ht]
\centering
\includegraphics[width=0.7\textwidth]{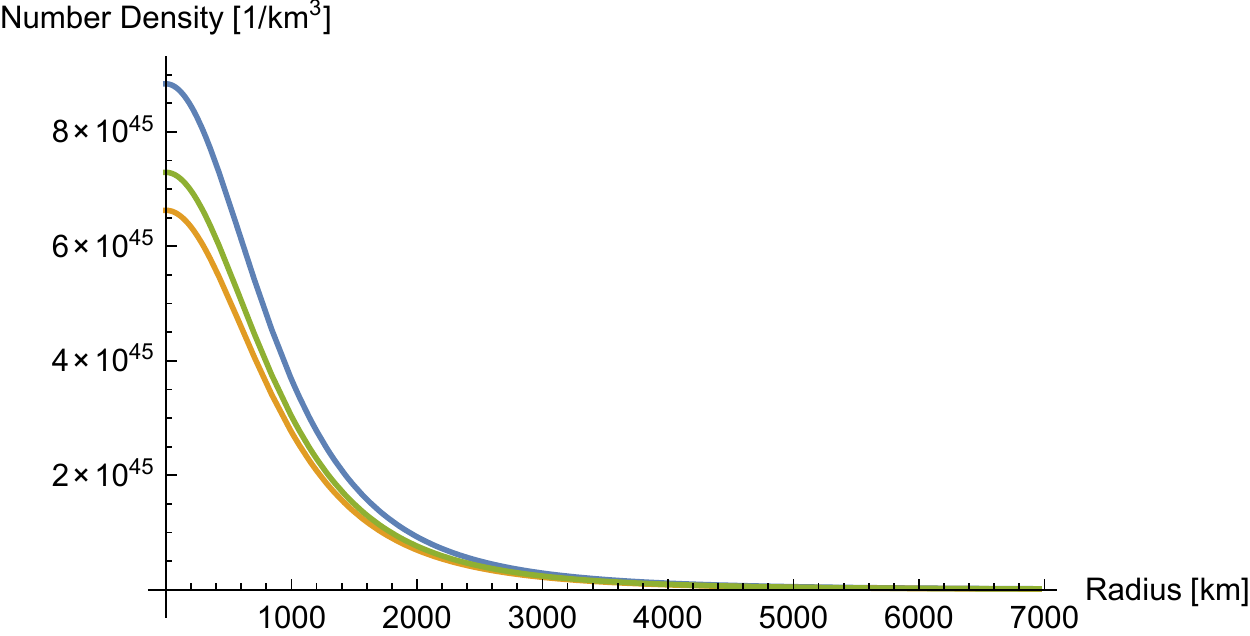}
 \caption{The number density in km$^{-3}$ as a function of radius for the prototypical WD used in our analysis ($R_*=0.0093R_{\odot}$ and $M_*=0.7M_{\odot}$). Blue curve: 100\% $^{12}$C, orange curve: 100\% $^{16}$O, green curve: 33\% $^{12}$C, 66\% $^{16}$O.}
       \label{wd_density}
\end{figure}

In our analysis, we need the number density of nuclei as a function of radius $n(r)$, so we assume that white dwarfs are modeled by polytropes of index 3 \cite{Araujo:2011hv,Chavanis:2006pf}. We choose the values of $R_*=0.0093R_{\odot}$ and $M_*=0.7M_{\odot}$ as the values of our prototypical white dwarf \cite{Hooper:2010es}. We show the number density of a white dwarf with these values of $R_*$ and $M_*$, assuming an $n=3$ polytrope, in Figure \ref{wd_density}. We display the number density assuming a white dwarf composed entirely of carbon-12, oxygen-16, and 33\% $^{12}$C, 66\% $^{16}$O. We assume that the elemental abundance does not depend on the radius. This is reasonable because both carbon and oxygen have similar scattering cross sections, and because the white dwarf is the core of a dead star \cite{Hurst:2014uda}.

\subsection{Capture rate in a white dwarf: geometric saturation}
\label{sec:geomrate}

The capture rate derived in the previous subsection is valid when scattering events are relatively rare. At large cross sections, however, it overestimates the total capture rate. The largest possible capture rate is a geometric one, when the star absorbs every dark matter particle that passes through its surface. This geometric rate is larger than the naive geometrical cross section of the star, $\sigma_0 = \pi R_*^2$, due to the long-range attractive force of gravity, which allows it to capture material from a wider region. For dark matter particles with velocity at infinity $v_\infty$, a simple estimate using conservation of energy and angular momentum gives a capture rate 
\be
\sigma = \pi b_{\rm max}^2 = \sigma_0 \left(1 + \frac{v_{\rm esc}^2}{v_\infty^2}\right),
\ee
where $v_{\rm esc}^2 = 2G_N M/R$ is the escape velocity at the stellar surface. The typical relative velocity of DM and the star will be of order the local velocity dispersion $v_0$. In time $t$ the star could swallow up all the dark matter in a volume of order $V = \sigma v_0 t$. For dark matter particles arising from a Maxwell-Boltzmann distribution with velocity dispersion $v_0 \ll v_{\rm esc}$, the associated geometric capture rate becomes \cite{Gould:1987ir,Hooper:2010es}
\begin{align}
C_{\rm geom} &= \sqrt{\frac{8\pi}{3}} \frac{3 G_N R_{\rm wd} M_{\rm wd} \rho_{\rm DM}}{m_{\rm DM} v_0} \nonumber \\
&\approx 7 \times 10^{24}~{\rm sec}^{-1} \frac{\rho_{\rm DM}}{0.3~{\rm GeV}/{\rm cm}^3} \frac{1000~{\rm GeV}}{m_{\rm DM}} \frac{R_{\rm wd}}{6.5 \times 10^8~{\rm cm}} \frac{M_{\rm wd}}{0.7 M_\odot} \frac{220~{\rm km}/{\rm sec}}{v_0}.
\label{eq:Cgeom}
\end{align}
Here we have used values of $\rho_{\rm DM}$ and $v_0$ typical for the region of the Milky Way in our vicinity. 

If dark matter annihilations have equilibrated with capture, then a corresponding conversion of dark matter mass to energy is taking place, corresponding to a luminosity
\be
L_{\rm max} = m_{\rm DM} C_{\rm geom} \approx 10^{25}~{\rm erg}/{\rm sec} \frac{\rho_{\rm DM}}{0.3~{\rm GeV}/{\rm cm}^3} \frac{R_{\rm wd}}{6.5 \times 10^8~{\rm cm}} \frac{M_{\rm wd}}{0.7 M_\odot} \frac{220~{\rm km}/{\rm sec}}{v_0},
\ee
using that $1~{\rm erg} \approx 624~{\rm GeV}$. Importantly, notice that this maximum possible luminosity is independent of the dark matter mass, depending only on the mass density of dark matter near the star. Measured white dwarf luminosities are in the range $10^{28}-10^{31}~{\rm erg/s}$ \cite{Bertone:2007ae}. This shows us that interesting results from white dwarf cooling are only accessible in regions of much larger dark matter density than our local neighborhood.

Importantly, the maximum luminosity increases with the dark matter density and is larger when the dark matter velocity dispersion $v_0$ is smaller:
\be
L_{\rm max} \propto \frac{\rho_{\rm DM}}{v_0}.
\ee
In the galactic center, $\rho_{\rm DM}$ is large but $v_0$ is not very small. This suggests that good targets are smaller dark matter halos, which can have high central densities and at the same time low velocity dispersions. We will return to this point in \S\ref{sec:observation}, after first arguing that our calculation of the capture rate for higgsinos really does correspond to an effective stellar luminosity.

\subsection{Aftermath of capture: de-excitation of the heavier higgsino}

The capture rate calculation we have presented so far simply determines the rate at which the process ${\widetilde H}^0_1 N \to {\widetilde H}^0_2 N$ produces a heavy mass eigenstate ${\widetilde H}^0_2$ with sufficiently low velocity that it cannot escape the star. We do not expect the heavy mass eigenstate to survive for a time comparable to the age of the star. Either it will downscatter back to the light mass eigenstate or it will decay. These two processes are kinematically distinct. The dominant decay in the range of mass splittings of interest is ${\widetilde H}^0_2 \to {\widetilde H}^0_1 \gamma$ (see \S\ref{sec:higgsino}). In the rest frame of the decaying heavier higgsino, the two decay products both acquire a small momentum of approximately $\delta$, the mass splitting. In the rest frame of the star, then, the daughter higgsino ${\widetilde H}^0_1$ has velocity approximately the same as the parent higgsino ${\widetilde H}^0_2$, and remains captured. On the other hand, the downscattering process ${\widetilde H}^0_2 N \to {\widetilde H}^0_1 N$ is exothermic, and if it is the dominant means of de-excitation we should be careful that it does not give enough energy to the daughter higgsino to undo the initial energy loss and eject the higgsino from the star.

We computed in equation (\ref{eq:decaywidth}) that for mass splittings of order an MeV, the decay length $c\tau$ of ${\widetilde H}^0_2$ is about $3 \times 10^7~{\rm cm}$; put differently, the particles decay in about a millisecond. At much smaller values of $\delta$ the lifetimes become significantly longer, but at small values of $\delta$, the inelasticity is not very important and ${\widetilde H}^0_1$ ejection after ${\widetilde H}^0_2$ capture is not a concern. So the question is what is the mean free time for ${\widetilde H}^0_2$ to downscatter. If we assume the ${\widetilde H}^0_2$ velocity is of order the escape velocity and compute $(n_{\rm C} \sigma_{\chi N} v_{\rm esc})^{-1}$ with the estimated cross section (\ref{eq:sigmachiN}) and the average number density of carbon atoms in the star, we obtain an estimated scattering time on the order of a millisecond, similar to the decay time. However, there is an additional kinematic fact that works in our favor. The carbon atoms in the star are typically moving quite slowly: the temperature of the star is far below the energy scale $\delta$. In the initial upscattering process, a ${\widetilde H}^0_1$ falling into the star picks up speed and hits a nucleus; part of its energy allows it to convert to ${\widetilde H}^0_2$, but to conserve momentum it gives the nucleus a kick. In fact, the outgoing nucleus typically has a velocity on the order of the escape velocity of the star or larger. The downscattering process cannot simply have the reverse kinematics, because there are no fast-moving nuclei on which to scatter. Because the downscattering process provides momentum to the escaping ${\widetilde H}^0_1$, it must also give a kick to the outgoing nucleus. In the regime of scattering where the ${\widetilde H}^0_1$ is ejected, the momentum transfer to the nucleus is large enough that the nuclear form factor suppresses the ejection rate. Thus, we believe that it is a reasonable approximation that the fate of the captured ${\widetilde H}^0_2$ is always to become a captured ${\widetilde H}^0_1$.

\subsection{Checking the equilibration of capture and annihilation}

The number of dark matter particles $N$ in the star evolves according to an equation
\be
dN/dt = C - C_A N^2,
\ee
where
\be
C_A = \frac{\int d^3 r\, n(r)^2 \left<\sigma_A v\right>}{\left(\int d^3 r\, n(r)\right)^2},
\ee
with $n(r)$ the number density of captured dark matter and $\left<\sigma_A v\right>$ the averaged annihilation cross section of the captured dark matter. The capture rate and annihilation rate will equilibrate, so that $dN/dt \approx 0$, after a time
\be
\tau_{\rm eq} \sim \frac{1}{\sqrt{C C_A}}.
\ee
We have computed $C$ and would like to interpret this as a measure of the luminosity, but to do so we should ensure that the stars we can observe have an age $t \gg \tau_{\rm eq}$. We will obtain a pessimistic estimate of $\tau_{\rm eq}$ by assuming that the number density is 
\be
n(r) = \frac{N}{R_{\rm star}^3},
\ee
so we can estimate
\be
C_A = \frac{1}{R_{\rm star}^3} \left<\sigma_A v\right> \approx 6 \times 10^{-53} {\rm sec}^{-1} \left(\frac{10^{-2}~R_\odot}{R_{\rm star}}\right)^3 \frac{\left<\sigma_A v\right>}{2 \times 10^{-26}~{\rm cm}^3/{\rm sec}},
\ee
where we have put in the thermal WIMP annihilation rate for $\left< \sigma_ A v\right>$ \cite{Steigman:2012nb}; the higgsino annihilation rate is of this order or larger in the mass range we are interested in.

This is a pessimistic estimate, because if the dark matter particles have scattered repeatedly in the star, they will thermally equilibrate and sink to the middle of the star, where their density will be much higher. Continuing with our pessimistic estimate, let us take $C = C_{\rm geom}$ from equation (\ref{eq:Cgeom}) of \S\ref{sec:geomrate}. Then we find an equilibration time
\be
\tau_{\rm eq} \sim \frac{1}{\sqrt{7 \times 10^{24} \times 6 \times 10^{-53}}}~{\rm sec} \approx 1.5 \times 10^6~{\rm yr}.
\ee
This million-year time scale is much less than the age of a typical white dwarf star. If we go toward lighter higgsinos or denser dark matter environments, both $C_{\rm geom}$ and $\left<\sigma_A v\right>$ increase, so the equilibration time becomes shorter. Hence, at least for white dwarf stars, we are usually safe assuming that the capture and annihilation rates have equilibrated.

Once the $\widetilde{H}^0_1$ particles have been captured and accumulate inside the star, they will annihilate dominantly to $W^+ W^-$ and $ZZ$ (with slightly more of the former). Both the $W$ and $Z$ decay dominantly to quarks, the decay products of which will hadronize and then rapidly thermalize in the medium of the star. A fraction of the decays will be to high-energy neutrinos that escape the star. In principle this makes compact stars point sources of high-energy neutrinos, but in practice the flux is too low to be observed by detectors like IceCube or Antares. Crudely, the dark matter annihilation rate in a volume $V$ is $\langle \sigma_A v \rangle n_{\rm DM}^2 V$. Comparing this to the annihilation rate in a star in the geometric regime, we see that a single white dwarf star has the same brightness in neutrinos as a region of space of volume
\be
V \sim \frac{G_N R_{\rm wd} M_{\rm wd}}{\langle \sigma_A v \rangle n_{\rm DM} v_0} \sim \left(0.03~{\rm pc}\right)^3 \frac{0.3~{\rm GeV}/{\rm cm}^3}{\rho_{\rm DM}} \frac{m_{\rm DM}}{1000~{\rm TeV}} \frac{220~{\rm km}/{\rm s}}{v_0}.
\ee
The inverse scaling with dark matter density is because annihilation in empty space requires two dark matter particles to meet, while the annihilation rate in the star (after equilibration) is set by the capture rate, which is linear in dark matter density. If the volume $V$ were overwhelmingly large compared to the typical volume in between white dwarfs in some region like the Galactic Center or a dwarf galaxy, it could improve their prospects as targets for dark matter searches with neutrino telescopes. (Currently, limits on dark matter annihilation to $W$ and $Z$ bosons from neutrino telescopes are much weaker than those from gamma ray telescopes.) If it were large in absolute terms, it would make white dwarfs interesting point sources. As it stands, the number is not large enough to make white dwarfs that have captured dark matter appealing targets for neutrino observations.

\subsection{Observational prospects}
\label{sec:observation}

Given that the neutrino flux is too small to be interesting, the best opportunity for observing the effect of higgsino capture in compact stars is through the heating of the star by annihilations, as discussed in \cite{Bertone:2007ae, Hooper:2010es, McCullough:2010ai, Fan:2011dw, Hurst:2014uda, Baryakhtar:2017dbj}. To put an upper bound on the density of higgsino dark matter, we would thus like to find old, cold white dwarfs, especially in regions of high dark matter density or low velocity dispersion. To make a {\em positive} case for the existence of higgsino dark matter, we would like to find evidence of a surprisingly high floor in the temperature of white dwarfs in a region of high dark matter density. These are challenging tasks.

\begin{figure}[!h]\begin{center}
\includegraphics[width=1.0\textwidth]{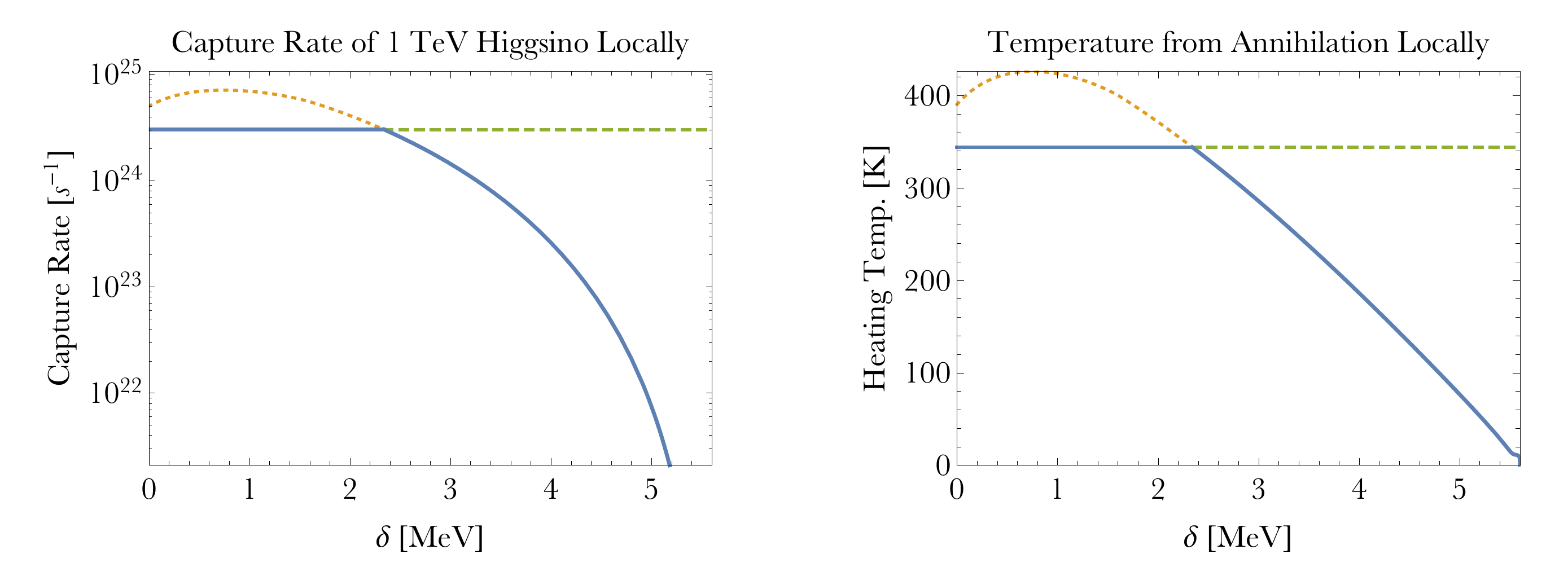}
\end{center}
\caption{Capture rate of 1 TeV higgsino dark matter in a white dwarf in our local neighborhood of the Milky Way (left), with the associated temperature derived from the luminosity associated with dark matter annihilations in equilibrium (right). The solid curve is the computed capture rate. The horizontal dashed line is the capture rate predicted at geometric saturation. The curved dotted line is the naive capture rate calculation neglecting geometric saturation.}
\label{fig:localcapture}
\end{figure}%

In Figure \ref{fig:localcapture} we display the capture rate for dark matter in white dwarf stars in our local neighborhood of the Milky Way, together with the temperature the star would have if the luminosity of the star is dominated by dark matter annihilations. In this estimate we have assumed that the dark matter annihilation does not substantially change the radius of the neutron star. This assumption is made throughout the literature on this topic, though revisiting it with a careful calculation would be worthwhile. We can see that these stars would be very cold---hundreds of Kelvins---and for this reason white dwarfs in our neighborhood are not good targets. For example, a white dwarf of mass $M_{\rm wd} = 1.05 M_\odot$ located about 267 parsecs away from us is constrained to have $T < 3000~{\rm K}$ \cite{Kaplan:2014mka}. It has not been directly observed; its existence is inferred from the motion of its pulsar companion. The limit on its temperature is not stringent enough to have implications for dark matter. Given that it inhabits our region of the Milky Way where $\rho_{\rm DM} \approx 0.3~{\rm GeV}/{\rm cm}^3$ and $v_0 \approx 220~{\rm km}/{\rm s}$, we predict a minimum luminosity from dark matter annihilations at geometric saturation of about $1.7 \times 10^{25}~{\rm erg}/{\rm s}$ while the temperature constraint implies $L \lesssim 3 \times 10^{28}~{\rm erg}/{\rm s}$. If future observations could improve the bound on the temperature of this white dwarf by an order of magnitude, there would be significant tension with expectations from dark matter capture and annihilation. However, white dwarfs are not expected to cool to hundreds of Kelvins over a time of order the age of the universe, so such a stringent temperature bound is unlikely. Other cool white dwarfs with temperatures in the $2000-3000~{\rm K}$ range have been identified as companions to millisecond pulsars located about a kiloparsec from us \cite{Testa:2015cha,Bassa:2015qga}. If similar observations could be made sufficiently close to the galactic center, where the dark matter density is larger, this could lead to either constraints on the dark matter capture rate or on the cuspiness of the dark matter distribution in the inner galaxy.

\begin{figure}[!h]\begin{center}
\includegraphics[width=0.8 \textwidth]{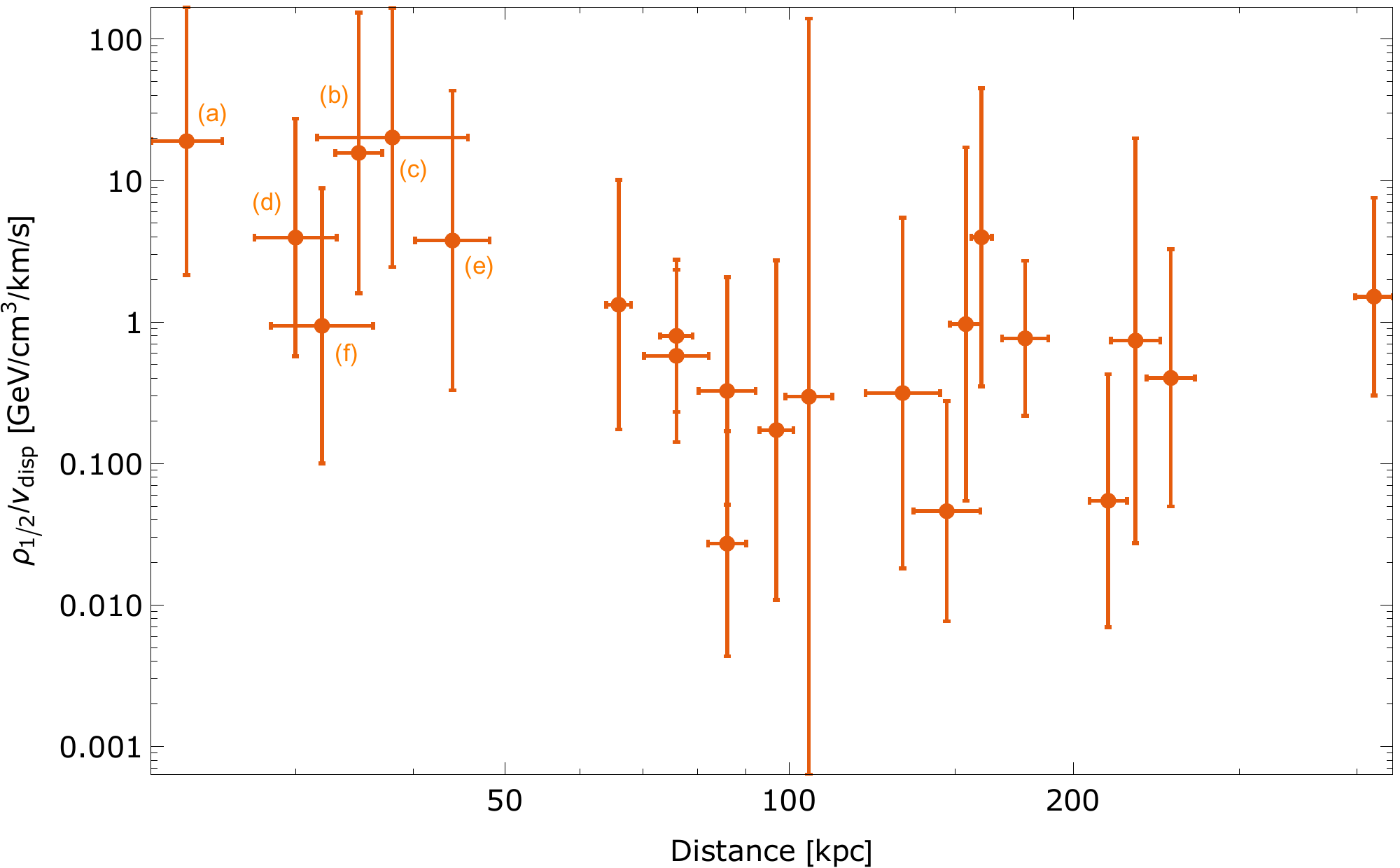}
\end{center}
\caption{Dwarf galaxies plotted by distance from us and their dark matter density at half-light radius $\rho_{1/2}$ divided by the velocity dispersion $v_{\rm disp}$. Six of the best targets for searches for dark matter capture in white dwarfs are labeled: (a) Segue I; (b) Segue II; (c) Willman 1; (d) Reticulum II; (e) Coma Berenices; (f) Ursa Major II.}
\label{fig:dwarfgalaxies}
\end{figure}

Since the luminosity from dark matter capture is proportional to the density of dark matter in the area of the white dwarf, and inversely proportional to the velocity dispersion, the best white dwarf candidates will be in  regions with a large value of $\rho_{\rm dm}/v_{\rm disp}$. Dwarf galaxies are a natural target. We display a number of dwarf galaxies sorted by $\rho_{\rm dm}/v_{\rm disp}$ and their distance from us in Figure \ref{fig:dwarfgalaxies}. To standardize the numbers in the plot, we use the dark matter density at the half-light radius. To find it, we find the NFW characteristic density and characteristic radii from $v_{\rm max}$ and $r_{\rm max}$, where $v_{\rm max}$ is the maximum circular velocity and $r_{\rm max}$ is the radius where this velocity is attained. We then use the NFW profile to find the density at the half-light radius. We use the values of $v_{\rm max}$ and $r_{\rm max}$ from \cite{Martinez:2013els}, with the exception of Reticulum II, whose values are obtained from \cite{Boddy:2017vpe}; Leo V, whose values are obtained from  \cite{Jiang:2015vra}; and the $v_{\rm max}$ for Sagittarius from \cite{Jiang:2015vra}. Then we find the NFW scale radii using $r_s=r_{\rm max}/2.163$ and the characteristic density with $\rho_s=(4.625/4\pi G)(v_{\rm max}/r_s)^2$. The half-light radii for the dwarf galaxies are obtained from \cite{McConnachie:2012vd}, with the exception of Leo T from \cite{Geringer-Sameth:2014yza} and Reticulum II from \cite{Simon:2015fdw}. Since the errors on $v_{\rm max}$ and $r_{\rm max}$ are asymmetric, to propagate errors, we use the average of the low and high errors. The distances and velocity dispersions are taken from \cite{McConnachie:2012vd}, with the exception of Reticulum II, whose distance and velocity dispersion are obtained from \cite{Simon:2015fdw}.

\begin{figure}[!h]\begin{center}
\includegraphics[width=1.0\textwidth]{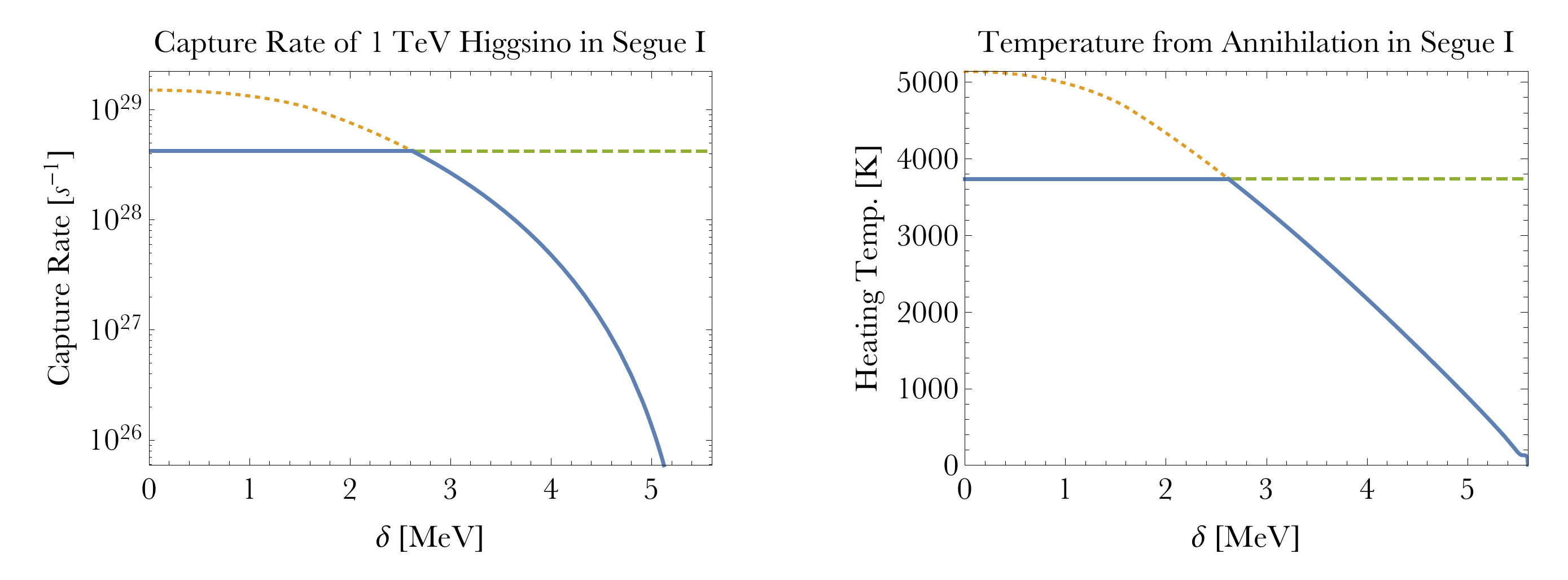}
\end{center}
\caption{Capture rate of 1 TeV higgsino dark matter in a white dwarf in Segue 1 (left), with the associated temperature derived from the luminosity associated with dark matter annihilations in equilibrium (right). The solid curve is the computed capture rate. The horizontal dashed line is the capture rate predicted at geometric saturation. The curved dotted line is the naive capture rate calculation neglecting geometric saturation.}
\label{fig:segue1capture}
\end{figure}

On the basis of Figure \ref{fig:dwarfgalaxies} we see that there are a handful of dwarf galaxies that have large $\rho_{\rm dm}/v_{\rm disp}$, a few orders of magnitude above that of our vicinity of the Milky Way, and are located within 50 kpc of us. Segue I, Segue II, Willman 1, and Reticulum II are among the best candidates. For example, Segue I is 23 kpc away, with a velocity dispersion of about $4~{\rm km}/{\rm sec}$ and estimated dark matter density at the half-light radius $70~{\rm GeV}/{\rm cm}^3$ (with large uncertainty). Notably, the best target dwarf galaxies have all been discovered within about the last decade, leaving some room for optimism that other so-far-undiscovered dwarf galaxies may also provide good targets. The capture rate for higgsino dark matter in a white dwarf in Segue I, which is about four orders of magnitude larger than the capture rate in our region of the galaxy, is illustrated in Figure \ref{fig:segue1capture}. The right panel shows that a floor on white dwarf temperatures of around 4000 K could be an indication of dark matter capture and annihilation. Such a temperature floor has been observed for white dwarfs in the globular cluster M4 \cite{Bedin:2009it}, which is located around 2 kpc from us. Segue I is ten times further away, and so observations of white dwarfs there are significantly more difficult.

Globular clusters like M4 have previously been suggested as a source of possible observational constraints on dark matter \cite{McCullough:2010ai, Hurst:2014uda}. However, globular clusters are not known to contain much dark matter, and are often thought to have formed in the disk of the galaxy rather than in a subhalo \cite{Hooper:2010es}. The exception may be certain globular clusters which began as dwarf galaxies and were tidally disrupted by the Milky Way. The globular cluster Omega Centauri ($\omega$ Cen) is an interesting candidate. Most of its dark matter content will have been tidally stripped, but unlike a true globular cluster is was formed in a dark matter halo and may still retain a significant amount of dark matter in its center \cite{Majewski:1999fw, Carraro:2000ji, Bekki:2003qw}. The cluster $\omega$ Cen is located 4.8 kpc from us (nearly a factor of 5 closer than Segue I), and has a central velocity dispersion of about $17~{\rm km}/{\rm s}$ \cite{Sollima:2009wh}. Because $\omega$ Cen is relatively nearby, the observational prospects for white dwarfs there are much better than in dwarf galaxies. Furthermore, the stellar mass of $\omega$ Cen is much larger, so there should be more white dwarfs to observe. Indeed, the cooling sequence in $\omega$ Cen has been studied \cite{Bellini:2013haa}. Its dark matter content is likely lower than that of Segue I, though estimates vary. For example, \cite{Amaro-Seoane:2015uny} postulates an enormous dark matter density in $\omega$ Cen due to an intermediate-mass black hole with an associated dark matter spike. However, even their baseline estimate of the dark matter density {\em without} the spike would already make $\omega$ Cen a better target than Segue I. Other estimates for the dark matter content remaining in $\omega$ Cen are lower \cite{Carraro:2000ji}, and there is no clear dynamical evidence for dark matter in the cluster \cite{DaCosta:2012wm}. As a result, the status of $\omega$ Cen as a probe of dark matter is currently rather unclear. Given that we already know a great deal about white dwarfs in the cluster, it would be important to find a strong argument---ideally empirical but perhaps based on simulations---for whether or not it should retain enough of a central dark matter density to host dark matter-burning white dwarf stars. This offers an interesting new opportunity for astrophysics to provide a window on particle physics.

\section{Conclusions and Outlook}
\label{sec:conclusions}

By surveying the current and near-future expected sensitivity of various experiments to higgsino dark matter, we have argued that this very simple WIMP candidate is likely to remain unconstrained in a window of small mass splittings (``pseudo-Dirac'' higgsinos) for some time to come. Currently indirect detection constraints set the most stringent bounds on this region of parameter space, with antiproton observations from AMS-02 leading the way albeit with substantial systematic uncertainties. The small mass splittings make observational tests that make crucial use of inelastic processes that can excite the heavier higgsino a promising possibility. We have discussed capture of higgsinos in white dwarf stars in some detail as an example of such a process. However, we find that deep observations of regions of high dark matter density and low velocity dispersion, such as dwarf galaxies, would be necessary to exploit this process. These will be challenging observations for astronomers to make, and so in the immediate future they are unlikely to surpass more traditional particle physics experiments in sensitivity. Still, in the longer term observations of dwarf galaxies as well as attempts to nail down the dark matter content of globular clusters (particularly those thought to have arisen from tidally disrupted dwarf galaxies) could offer a great deal of insight into particle physics.

It is worth considering whether inelastic processes open up any other novel means of looking for dark matter. One possibility is to exploit the fact that typical velocity dispersions in galaxy clusters are on the order of $1000-2000~{\rm km}/{\rm s}$, so that a TeV dark matter particle has kinetic energy around $6-20~{\rm MeV}$. This means that a process like ${\widetilde H}^0_1 {\widetilde H}^0_1 \to {\widetilde H}^0_2 {\widetilde H}^0_2$ can occur at a potentially interesting rate, following which the heavier ${\widetilde H}^0_2$ particle will decay to its lighter partner plus a monochromatic photon of energy $\delta$. Currently observational constraints on MeV gamma rays range are rather weak, but future missions like e-ASTROGAM will provide better coverage of this energy range \cite{DeAngelis:2016slk}. However, it appears unlikely that the constraints will be able to probe higgsino dark matter. As a quick back-of-the-envelope assessment, note that observations of the galactic center are expected to probe $\chi \chi \to \gamma \gamma$ cross sections $\langle \sigma v\rangle_A \sim 10^{-34}~{\rm cm}^3/{\rm s}$ for $m_\chi \approx 1~{\rm MeV}$ \cite{Bartels:2017dpb}. The higgsino upscattering process would also lead to MeV gamma rays, but the number density of TeV higgsino dark matter particles is smaller than that of MeV dark matter particles by a factor of $10^{-6}$, so the ``effective'' $\langle \sigma v\rangle_A$ that could be observed is smaller by 12 orders of magnitude. This means that we would need a rate for ${\widetilde H}^0_1 {\widetilde H}^0_1 \to {\widetilde H}^0_2 {\widetilde H}^0_2$ of order $10^{-22}~{\rm cm}^3/{\rm s}$ to match a $10^{-34}~{\rm cm}^3/{\rm s}$ $\chi \chi \to \gamma\gamma$ observation. The higgsino cross section is set by the weak scale and suppressed by the small final-state phase space, and falls short of this target. Furthermore, \cite{Bartels:2017dpb} considered observations of the galactic center, but higgsinos in the galactic center would lack sufficient energy for this process, and galaxy clusters would be the right target. But in general observations of gamma rays in galaxy clusters have lagged behind observations of the galactic center in sensitivity to annihilating dark matter. On the basis of these back-of-the-envelope considerations, the higgsino excitation followed by decay process appears to have a rate at least a few orders of magnitude too small to be within range of e-ASTROGAM. 

While a combination of collider searches, direct detection, and indirect detection has now ruled out huge swaths of the parameter space for WIMP dark matter, especially in the case of WIMPs with Standard Model electroweak gauge interactions, we have seen that the last electroweak WIMP, the pseudo-Dirac higgsino, is still healthy. Better modeling of cosmic ray propagation in the galaxy, which could improve our understanding of antiprotons observed by experiments like AMS-02, is one route by which astrophysics could help close the book on the last WIMP. A better understanding of the dark matter content of globular clusters is another. These offer exciting opportunities at the intersection of astronomical observation and theoretical modeling that could have huge ramifications for our understanding of particle physics.

\section*{Acknowledgments}

We thank Prateek Agrawal, Matt Baumgart, JiJi Fan, Liam Fitzpatrick, and Wei Xue for useful conversations. MR is supported in part by the NSF Grant PHY-1415548 and in part by the NASA ATP Grant NNX16AI12G.

\bibliography{ref}
\bibliographystyle{utphys}

\end{document}